\documentclass[journal,twoside,web]{ieeecolor}
\usepackage{xcolor}
\usepackage{generic}
\usepackage{etoolbox}
\patchcmd{\thebibliography}
  {\section*{\color{black}}}
  {\section*{\color{black}References}}
  {}
  {\PackageWarning{bibliography}{Could not restore References heading}}
\definecolor{nblue}{rgb}{0,0.263,0.576}
\definecolor{mblue}{rgb}{0.075,0.541,0.855}
\usepackage{amsmath,amsfonts,amssymb}  
\usepackage{mathtools}
\usepackage{graphicx}
\usepackage{epsfig}
\usepackage{accents}
\usepackage{booktabs}

\usepackage{cite}
\usepackage{hyperref}
\hypersetup{hidelinks}
\usepackage{textcomp}
\usepackage{balance}
\usepackage{picinpar}
\usepackage{url}
\usepackage{flushend}
\usepackage{colortbl}
\usepackage{placeins}
\usepackage{xcolor}
\usepackage{svg}
\usepackage{soul}
\usepackage{multirow}
\usepackage{tikz}
\usetikzlibrary{shapes,shapes.geometric,arrows,fit,calc,positioning,automata}
\usepackage{verbatim}
\usepackage{pifont}
\usepackage{alltt}
\usepackage{siunitx}
\makeatletter
\def\endfigure{\end@float}
\def\endtable{\end@float}
\makeatother
\usepackage[caption=false,font=footnotesize]{subfig}
\usepackage{epstopdf}
\usepackage{pbox}
\usepackage{float}
\usepackage{lipsum}
\let\IEEEproof\proof
\let\endIEEEproof\endproof
\let\proof\relax
\let\endproof\relax
\usetikzlibrary{calc,arrows.meta,backgrounds,fit}
\definecolor{leadcol}{HTML}{F26B3A}   
\definecolor{follcol}{HTML}{2563EB}   
\definecolor{roadcol}{HTML}{2D3239}
\usepackage{amsthm}
\let\proof\IEEEproof
\let\endproof\endIEEEproof
\theoremstyle{plain}
\newtheorem{thm}{Theorem}

\theoremstyle{definition}
\newtheorem{assum}{Assumption}

\theoremstyle{rem}
\newtheorem{rem}{Remark}
\usepackage{orcidlink}
\let\labelindent\relax
\usepackage
{
    graphicx,
    dsfont, 
    xcolor,
    mathtools,
    authblk,
    enumitem,
    nicefrac,
    array,
}

\usepackage[utf8]{inputenc}
\usepackage{dblfloatfix}

\graphicspath{{Figures/}}

\newcommand{\R}{\mathbb{R}}                                     

\newcommand{\N}{{\mathbb{N}}}                                   

\newcolumntype{C}[1]{>{\centering\let\newline\\\arraybackslash\hspace{0pt}}m{#1}}

\newcommand{\vmax}{v_{\max}}

\DeclareMathOperator{\sgn}{sgn}
\DeclareMathOperator{\erf}{erf}

\newcommand{\cS}{\mathcal{S}}

\DeclareOldFontCommand{\rm}{\normalfont\rmfamily}{\mathrm}
\def\BibTeX{{\rm B\kern-.05em{\sc i\kern-.025em b}\kern-.08em
    T\kern-.1667em\lower.7ex\hbox{E}\kern-.125emX}}
\begin{document}
\title{Safe Platooning under Input Constraints with Link-Wise Velocity String Stability}
\author{Panagiotis S. Trakas, Charalampos P. Bechlioulis, and Thomas Berger
\thanks{T. Berger acknowledges funding by the Deutsche Forschungsgemeinschaft (DFG, German Research Foundation) - Project-ID 544702565.}
\thanks{
P. S. Trakas is with the Department of Information Technology and Electrical Engineering, ETH Zurich, Switzerland. (e-mail: ptrakas@control.ee.ethz.ch)}
\thanks{C. P. Bechlioulis is with the Department of Electrical and Computer Engineering, University of Patras, Greece. (e-mail: chmpechl@upatras.gr)}
\thanks{T. Berger is with the Martin Luther University Halle-Wittenberg, Institute for Mathematics, Theodor-Lieser-Str. 5, 06120 Halle (Saale), Germany. (e-mail: thomas.berger@mathematik.uni-halle.de)}
}
\maketitle
\begin{abstract}
We propose a decentralized controller for vehicle platoons with different braking and propulsion limits. The method uses only local distance and velocity measurements and requires neither online optimization nor exact vehicle parameters. The design combines a braking-distance safety margin with an adaptive funnel that relaxes the spacing requirement during propulsion saturation. A standstill mechanism prevents backward motion and supports stop-and-go operation. Under explicit uniform conditions on the vehicle parameters, controller settings, and initial states, we prove that closed-loop solutions exist for all time, collisions are avoided, and input limits are satisfied. Moreover, each follower tracks its predecessor’s velocity range with gain one, up to an explicit additive offset determined by the funnel width and the braking-distance term. This bound establishes link-wise velocity string stability. Both the offset and the bounds on speeds and spacing errors are independent of vehicle index and platoon length. Simulations with 2 to 100 followers illustrate safety and bounded velocity deviations under disturbances and actuator limits.
\end{abstract}

\begin{IEEEkeywords}
Autonomous driving; decentralized control; funnel control; string stability;
vehicle platoons; input constraints.
\end{IEEEkeywords}

\section{Introduction}\label{sec:introduction}
\IEEEPARstart{V}{ehicle} platooning can increase road capacity and reduce energy consumption by allowing vehicles to travel safely with smaller gaps between them~\cite{AlamEtAl15}.
Close spacing, however, requires each follower to respond promptly to speed changes in the vehicle ahead.
When the leader brakes, every follower must avoid a collision while respecting its braking limits.
At the same time, speed changes should not grow as they pass from one vehicle to the next.
Experiments with commercially available adaptive cruise control systems show that this amplification can occur in practice~\cite{GuntEtAl21}.

String stability expresses the requirement that disturbances should not grow as they pass from one vehicle to the next~\cite{SwarHedr96,FengEtAl19,BessJoha17}.
For this guarantee to remain useful in long platoons, the bound between consecutive vehicles must not depend on their position or the total number of vehicles.
This is difficult when each follower uses information only from the vehicle directly ahead, especially when the desired gap is constant~\cite{SeilPantHedr04,WijnBess20}.
Cooperative adaptive cruise control can help reduce disturbance amplification by sharing information from the leader or several vehicles ahead~\cite{NausEtAl10,OncuEtAl14}.
However, these methods depend on communication and can be affected by delays, packet loss, and changes in the communication links.  In this work, we instead seek a
decentralized controller that uses only onboard measurements of distance and velocity.

String stability alone does not guarantee collision avoidance.
A platoon may reduce small speed disturbances but still become unsafe during hard braking.
Each vehicle can apply only a limited braking force, regardless of what the controller requests.
Braking and propulsion limits must therefore be included in the safety analysis.
Simply applying saturation to a controller designed without these limits can invalidate the conditions used to prove collision avoidance.

Several control methods address different aspects of this problem.
Control barrier function methods enforce conditions for collision avoidance by solving quadratic programs online~\cite{AmesEtAl14,AmesEtAl17}.
Model predictive control can enforce safety and input limits while preserving string stability in the nominal system~\cite{KianFalcFred15}.
These approaches handle constraints directly, but require online optimization and sufficiently accurate models.
Nonlinear cruise controllers based on barrier arguments guarantee collision avoidance, positive velocities, bounded acceleration, and string stability under suitable restrictions on the feedback laws~\cite{KaraTheoPapa23}.

Actuator saturation has also been studied directly in platoon control. Tube-based methods account for disturbances and saturation in discrete-time models~\cite{FengEtAl20}, while variable-headway designs address input saturation in connected platoons~\cite{ChenEtAl21}. The constraint-oriented controller in~\cite{YangEtAl20} uses bidirectional vehicle information and anti-windup compensation to guarantee bounded spacing errors and collision avoidance. Prescribed-performance designs have also been developed for large nonlinear platoons~\cite{VergEtAl18}, including controllers that handle actuator dead zones and saturation through smooth approximations and adaptive compensation~\cite{GaoZhangGuo23}. These works provide important solutions, but they use different communication structures, vehicle models, and notions of string stability. However, by themselves quantify how velocity variations propagate from one vehicle to the next under the actuator
limits.

Our work builds on the decentralized funnel controller in~\cite{BergBess26} for heterogeneous nonlinear platoons. That work uses only local measurements and guarantees collision avoidance and practical velocity string stability.
However, its safety proof assumes that the requested control force can always be applied and does not account for braking and propulsion constraints.
During emergency braking, the controller may request more braking force than a vehicle can provide, so the safety guarantee may no longer apply. Furthermore, during acceleration, insufficient propulsion may cause the spacing error to approach the funnel boundary, where the controller becomes singular and the requested control signal becomes unbounded.

Input-constrained funnel and adaptive performance controllers have been studied for general nonlinear systems~\cite{Berg24,TrakBech23,TrakBech24,Berg26}. In particular, the adaptive cruise controller in~\cite{acc25} guarantees collision avoidance and enforces amplitude and rate limits for a single follower. A single follower, however, does not capture disturbance propagation through a platoon and cannot provide a bound that is uniform in the platoon length. Our aim is to extend the local sensing approach of~\cite{BergBess26} to vehicles with different braking and propulsion limits, while guaranteeing collision avoidance and link-wise velocity string stability.

The main contributions are as follows.
\begin{enumerate}[label=(C\arabic*),leftmargin=*]
\item We propose a decentralized controller for heterogeneous vehicles with different braking and propulsion limits. Safety is built around a relative braking-distance term computed from the worst-case road grade, disturbance, vehicle mass, and uniformly available braking authority. Thus, the safety margin remains valid even when full braking is applied. A dynamically adjusted funnel boundary relaxes the spacing requirement during sustained propulsion saturation and contracts again when sufficient control authority becomes available. A standstill mechanism prevents backward motion and enables stop-and-go operation.

\item Under explicit uniform conditions on the vehicle parameters, controller settings, and initial conditions, we prove global existence, collision avoidance, satisfaction of every actuator constraint and link-wise velocity string stability. The bounds on the speeds, the spacing errors, and the controller signals do not depend on the position of a vehicle in the platoon or on the total number of vehicles. Moreover, each follower tracks the velocity range of its predecessor with gain one up to an explicit offset, which is independent of the vehicle index and the platoon length and consists of the funnel width, converted into velocity by the time headway, and of the braking distance released when the predecessor accelerates.

\item We evaluate the controller in stop-and-go simulations with 2 to 100 followers with different masses and actuator limits.
The simulations illustrate collision avoidance under disturbances, funnel adaptation during sustained propulsion saturation, and sequential stopping and restarting. For the larger platoons tested, the maximum velocity deviation from the leader remains nearly unchanged, while the response at the rear becomes more delayed.
\end{enumerate}

\subsection{Nomenclature}

In the following let $\N$ denote the natural numbers, $\N_0 = \N \cup\{0\}$, and $\R_{\ge 0} =[0,\infty)$. By $\|x\|$ we denote the Euclidean norm of $x\in\R^n$. For some interval $I\subseteq\R$, some $V\subseteq\R^m$ and $k\in\N$, $L^\infty(I, \R^{n})$ is the Lebesgue space of measurable, essentially bounded functions $f\colon I\to\R^n$, $W^{k,\infty}(I, \R^{n})$ is the Sobolev space of all functions $f:I\to\R^n$ with $k$-th order weak derivative $f^{(k)}$ and $f,f^{(1)},\ldots,f^{(k)}\in L^\infty(I, \R^{n})$, and $C^k(V,  \R^{n})$ is the set of $k$-times continuously differentiable functions $f: V \to \R^{n}$, with $C(V,  \R^{n}) := C^0(V, \R^{n})$.

\subsection{Vehicle dynamics}

\begin{figure*}
\centering
\resizebox{\textwidth}{!}{%
\begin{tikzpicture}[
    >={Stealth[length=2.4mm,width=2mm]},
    tag/.style={fill=#1, text=white, rounded corners=2.2mm,
                inner xsep=2.6mm, inner ysep=1.1mm,
                font=\small\sffamily\bfseries},
    panel/.style={fill=#1!8, draw=#1!55, line width=0.5pt,
                  rounded corners=3.5mm, inner sep=2.2mm},
    vel/.style={-{Stealth}, line width=1.5pt, color=#1},
    posarrow/.style={{Bar[width=2.6mm]}-{Stealth}, line width=0.9pt, color=#1},
    guide/.style={dotted, line width=0.9pt, color=#1!70},
    lbl/.style={font=\normalsize, inner sep=1pt}
]

\newcommand{\car}{\scalebox{-1}[1]{\includegraphics[width=34mm]{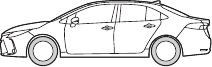}}}
\node (vehN) at (-90mm,0mm) [inner sep=0mm] {\car};
\node (veh1) at (-30mm,0mm) [inner sep=0mm] {\car};
\node (veh0) at ( 24mm,0mm) [inner sep=0mm] {\car};

\node at ($(vehN.east)!0.5!(veh1.west)$) [xshift=4mm] {\Large$\cdots$};

\begin{scope}[on background layer]
  \node[panel=follcol, fit=(vehN)] (pN) {};
  \node[panel=follcol, fit=(veh1)] (p1) {};
  \node[panel=leadcol, fit=(veh0)] (p0) {};
\end{scope}
\node[tag=follcol] at (pN.north) {Vehicle $N$};
\node[tag=follcol] at (p1.north) {Vehicle $1$};
\node[tag=leadcol] at (p0.north) {Leader};

\coordinate (roadL) at ($(vehN.south west)+(-18mm,0)$);
\coordinate (roadR) at ($(veh0.south east)+(18mm,0)$);
\coordinate (rb)    at ($(vehN.south)+(0,-3.4mm)$);          
\fill[roadcol, rounded corners=1.4pt] (roadL) rectangle (roadR |- rb);
\draw[white!85!roadcol, line width=0.7pt, dash pattern=on 4mm off 3mm]
     ($(roadL)+(2mm,-1.7mm)$) -- ($(roadR)+(-2mm,-1.7mm)$);

\draw[vel=follcol] (pN.east) -- +(10mm,0) node[midway, above=0.5mm, lbl, text=follcol] {$v_N(t)$};
\draw[vel=follcol] (p1.east) -- +(10mm,0) node[midway, above=0.5mm, lbl, text=follcol] {$v_1(t)$};
\draw[vel=leadcol] (p0.east) -- +(10mm,0) node[midway, above=0.5mm, lbl, text=leadcol] {$v_0(t)$};

\coordinate (O)  at ($(vehN.south west)+(-12mm,0)$);          
\coordinate (yA) at ($(vehN.south)+(0,-9mm)$);
\coordinate (yB) at ($(vehN.south)+(0,-15mm)$);
\coordinate (yC) at ($(vehN.south)+(0,-21mm)$);

\draw[guide=follcol] (vehN.south west |- rb) -- (vehN.south west |- yA);
\draw[guide=follcol] (veh1.south west |- rb) -- (veh1.south west |- yB);
\draw[guide=leadcol] (veh0.south west |- rb) -- (veh0.south west |- yC);

\draw[posarrow=follcol] (O |- yA) -- (vehN.south west |- yA)
     node[midway, above=0.3mm, lbl, text=follcol] {$x_N(t)$};
\draw[posarrow=follcol] (O |- yB) -- (veh1.south west |- yB)
     node[midway, above=0.3mm, lbl, text=follcol] {$x_1(t)$};
\draw[posarrow=leadcol] (O |- yC) -- (veh0.south west |- yC)
     node[pos=0.7, above=0.3mm, lbl, text=leadcol] {$x_0(t)$};

\end{tikzpicture}
}
\caption{Framework for the control of a vehicle platoon}
\label{Fig:platoon}
\end{figure*}
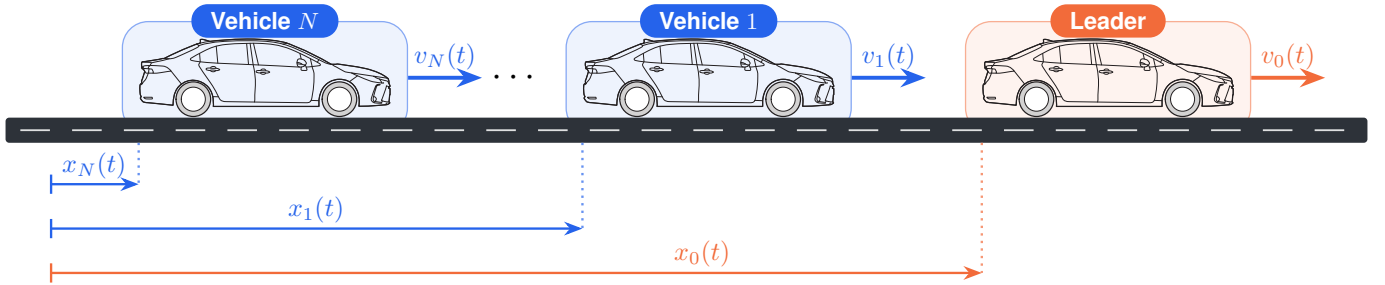

Consider a platoon of $N$ heterogeneous vehicles\footnote{In this work, the leader vehicle is not counted as a member of the platoon, it assumes a special role.} with nonlinear dynamics \cite{aastrom}:
\begin{equation}\label{eq:Sys}
\begin{aligned}
\dot x_i(t) &= v_i(t),\\
m_i\dot v_i(t)
&=\varphi_i\bigl(t,x_i(t),v_i(t),u_i(t)\bigr)\\
&:=u_i(t)-f_i\bigl(t,x_i(t),v_i(t)\bigr)+d_i(t).
\end{aligned}
\end{equation}
with $i=1,\ldots,N$, where $x_i(t)\in\R$ and $v_i(t)\in\R$ denote the position and velocity, respectively, of vehicle $i$ at time $t\geq0$, see also Fig.~\ref{Fig:platoon}. Moreover, $m_i$ (in \si{\kilo\gram}) is the mass of vehicle~$i$, $d_i\in L^\infty(\R_{\ge 0},\R)$ is a bounded disturbance (capturing modelling errors, uncertainties and noises), and the nonlinear function $f_i$ is the sum of the forces due to gravity~$F_{i,g}$, the aerodynamic drag~$F_{i,a}$ and the rolling friction~$F_{i,r}$, that is
\[
    f_i(t,x,v) = F_{i,g}(x) + F_{i,a}(t,x,v) + F_{i,r}(v),
\]
where, to keep the model smooth, the discontinuous sign function usually appearing in the rolling friction term is replaced by the smooth approximation $s_\kappa(v):=\erf(\kappa v)$, with
$\erf(z)=\frac{2}{\sqrt{\pi}}\int_0^z e^{-t^2}\,dt$ and a fixed (large) $\kappa>0$,
satisfying $\lim_{\kappa\to\infty}s_\kappa(v)=\sgn(v)$. The component forces are
\begin{align*}
  F_{i,g}:&\ \R\to\R,\ x\mapsto m_i\, g \sin \theta_i(x),\\
  F_{i,a}:&\ \R_{\ge 0}\times \R\times \R\to\R,\\
  & (t,x,v)\mapsto \tfrac12 \rho_i(t,x) C_{i,d}\, A_i \,s_\kappa(v)\, v^2,\\
  F_{i,r}:&\ \R\to\R,\ v\mapsto m_i\, g\, C_{i,r} \,s_\kappa(v).
\end{align*}
Here, $g = \SI{9.81}{\metre\per\second^2}$ is the acceleration of gravity,~$\theta_i(x)\in [-\tfrac{\pi}{2}, \tfrac{\pi}{2}]$ (in \si{\radian}) and $\rho_i(t,x)$ (in \si{\kilo\gram\per\metre^3}) denote the slope of the road and the (bounded) density of air at time~$t$ and location~$x$ for vehicle~$i$, resp.,~$C_{i,d}$ denotes the (dimensionless) shape-dependent aerodynamic drag coefficient,~$C_{i,r}$ the (dimensionless) coefficient of rolling friction, and~$A_i$ (in \si{\metre^2}) the frontal area of vehicle~$i$, respectively. The control input $u_i$ is the longitudinal force generated by propulsion or braking and transmitted through the tires. Positive values correspond to propulsion and negative values to braking.

The initial conditions for~\eqref{eq:Sys} and $i=1,\ldots,N$, are
\begin{equation}\label{eq:IC}
  x_i(0) = x_i^0\in\R,\ \ v_i(0) = v_i^0\in\R,\ \ x_i^0 < x_{i-1}^0
\end{equation}
where $x_0^0 = x_0(0)\in\R$ is the initial position of the leader vehicle which has a position trajectory $x_0 \in C^1(\R_{\ge 0},\R)$ and we set $v_0 := \dot x_0$.

In the following, we make precise the assumptions on the system parameters required in this work, which are standard and can be satisfied in any platoon.

\begin{assum}\label{ass:leader}
The leader position trajectory $x_0 \in C^1(\R_{\ge 0},\R)$ is such that $v_0 := \dot x_0$ is Lipschitz continuous with constant $\bar a_0>0$, and satisfies $0 \le v_0(t) \le \bar v_0$ for all $t \ge 0$.
\end{assum}

\begin{assum}\label{ass:bounds}
There exist known positive constants $\bar c_\rho$, $\underline c_\rho$,
$\bar d$, $\bar C_r$, $\underline C_r$, $\underline m$, $\bar m$ and a constant
$\bar \theta \in [0,\tfrac{\pi}{2})$, such that for all $i=1,\dots,N$, almost all $t\ge 0$ and all $x\in\R$ we have:
\[
\underline c_\rho \le \tfrac12 \rho_i(t,x) C_{i,d}\, A_i \le \bar c_\rho,
\quad
|\theta_i(x)| \le \bar \theta,
\quad
|d_i(t)| \le \bar d,
\]
\[
\underline m \le m_i \le \bar m,
\quad
\underline C_r \le C_{i,r} \le \bar C_r.
\]
Moreover, $\theta_i$ and $\rho_i$ are locally Lipschitz in $x$, uniformly on
compact time intervals, and measurable in $t$; together with the smoothness of
$s_\kappa$, this renders $f_i$ locally Lipschitz in the state, uniformly on compact
time intervals, and measurable in $t$.
\end{assum}

\subsection{Control objectives}\label{Ssec:ContrObj}

The control objective is to
\begin{enumerate}[label=(O\arabic*),leftmargin=*]
  \item guarantee collision avoidance between any two vehicles,
  \item satisfy prescribed input constraints for each vehicle,
  \item achieve string stability of the controlled platoon,
  \item only use decentralized controllers based on local information.
\end{enumerate}

We like to note at this point that objective (O2) above differs from our previous work~\cite{BergBess26}, where it was formulated as ``ensure a good traffic flow''. Instead, in the present paper we focus on satisfying prescribed input constraints, which do not simultaneously allow to control the traffic flow by prescribed upper bounds on the vehicle distances.

Clearly, (O1) can be formalized as 
\[
    \forall\, i=1,\ldots,N\ \forall\, t\ge 0:\  x_{i}(t) - x_{i-1}(t) < 0.
\]
In order to formalize~(O2), each vehicle is subject to local actuator constraints of the form
\begin{equation}\label{eq:actuator_constraints}
\underline u_i \le u_i(t) \le \bar u_i, \qquad t\ge 0
\end{equation}
for $i=1,\ldots,N$, where the bounds are given by
\begin{equation}\label{eq:control_bounds}
\underline u_i=-c_{d,i}m_i g,
\qquad
\bar u_i=c_{a,i}m_i g
\end{equation}
and the \textit{known} constants $c_{d,i},c_{a,i}>0$ represent the normalized
braking and propulsion limits of vehicle~$i$, respectively. We assume
that there exist known constants
$\underline c_d,\bar c_d,\underline c_a,\bar c_a>0$, independent of
$i$ and~$N$, such that
\[
\underline c_d\le c_{d,i}\le\bar c_d,
\qquad
\underline c_a\le c_{a,i}\le\bar c_a,
\qquad i=1,\ldots,N.
\]
The dimensionless parameters $c_{d,i}$ and $c_{a,i}$ express the
vehicle-specific braking and propulsion limits as multiples of $m_i g$.
They may be selected according to the vehicle specifications or chosen
more conservatively to impose ride-comfort limits. For reference,
$c_{d,i}=0.8$ and $c_{a,i}=0.5$ correspond to nominal input-force
acceleration equivalents of $-0.8g$ and $0.5g$, respectively, before
road grade, resistance forces, and disturbances are taken into account.

For string stability as in~(O3) we view the velocity profile~$v_0$ of the leader vehicle as the disturbance/excitation of the platoon and require, in the spirit of $L^\infty$-string stability~\cite{SwarHedr96,BessJoha17}, that each follower tracks the velocity range of its
predecessor with gain one, up to an additive offset which is independent of the vehicle index and of the platoon length: there exists $\Theta>0$ such that for all $N\in\N$, all $i=1,\dots,N$, all admissible $v_0$ and all $t\ge0$ we have
\begin{equation}\label{eq:string-stab}
\begin{aligned}
  &\min\Big\{\min_{s\in[0,t]}v_{i-1}(s),v_i(0)\Big\}-\Theta  \\
  &\qquad\qquad \le v_i(t)\le
  \max\Big\{\max_{s\in[0,t]}v_{i-1}(s),v_i(0)\Big\}+\Theta .
\end{aligned}
\end{equation}
 Here~$v_i$ denotes the solution of~\eqref{eq:Sys} under an appropriate feedback~$u_i$ and for relevant classes of initial conditions and external disturbances, all of  which will be defined below. We call~\eqref{eq:string-stab} \emph{link-wise velocity string stability}. It differs
from the practical velocity string stability considered in~\cite{BergBess26}, which concerns the
velocity deviation from the leader, in that it quantifies the propagation of velocity
variations from each vehicle to its follower. A leader-relative bound follows from~\eqref{eq:string-stab}
by induction over $i$, which implies
\begin{equation*}
  \forall\,t\ge0:\ |v_i(t)-v_0(t)|\le
  \Big(\sup_{s\ge0}v_0(s)-\inf_{s\ge0}v_0(s)\Big)+I_i+i\,\Theta,
\end{equation*}
where $I_i\ge0$ measures the mismatch of the initial velocities $v_1(0),\dots,v_i(0)$
with the range of $v_0$ and vanishes if all of them lie in this range. The deviation
from the leader is thus bounded by the velocity oscillation of the leader up to an
offset that accumulates along the platoon. An unbounded accumulation must be additionally prevented by the control.

To achieve~(O4), in Section~\ref{Sec:ContrStruc} we present a
decentralized control design which, for vehicle~$i$, requires
instantaneous measurements of the inter-vehicle displacement
$x_i(t)-x_{i-1}(t)$, the ego velocity $v_i(t)$, and the relative
velocity $\Delta v_i(t):=v_i(t)-v_{i-1}(t)$. These quantities are
locally available from onboard velocity and ranging sensors, such as
forward-looking radar. In particular, the relative squared velocity
required in~\eqref{eq:db} can be computed as
\[
v_i(t)^2-v_{i-1}(t)^2
=2v_i(t)\Delta v_i(t)-\Delta v_i(t)^2.
\]
If measurement of the relative velocity is unavailable, then the predecessor velocity may conservatively be
set to zero in~\eqref{eq:db}, since $v_{i-1}(t)^2\ge0$, at the expense
of a larger braking-distance term. Apart from these measurements, the
controller requires no exact knowledge of the system parameters or
initial values.

\subsection{Organization of the Paper}

The remainder of the paper is organized as follows.
Section~\ref{Sec:ContrStruc} introduces the decentralized high-level
controller and the low-level standstill mechanism.
Section~\ref{stability} establishes global existence, collision
avoidance, uniform boundedness of the closed-loop signals, and link-wise velocity
string stability under the heterogeneous input constraints.
Section~\ref{Sec:Sim} presents numerical simulations, and
Section~\ref{Sec:Concl} concludes the paper.

\section{Decentralized control design}\label{Sec:ContrStruc}

In this section, we introduce a controller aimed at achieving the control objectives (O1)--(O4). We start with a low-level mechanism which prevents reverse motion, followed by considerations on the braking distance which are required to meet the input constraints, and finally the high-level controller is presented.

\subsection{Low-level Standstill mechanism}

Since reverse motion is outside the present platooning problem, each follower is
equipped with a low-level standstill mechanism, such as brake-hold or hill-hold, to prevent backward motion at zero velocity.
The mechanism releases the vehicle when the net longitudinal force $\varphi_i$ becomes positive, allowing forward motion. Similar lower-velocity-bound conventions are standard in full-range adaptive cruise control, e.g., \cite{althoff} excludes backward driving by imposing a zero lower velocity bound
at the vehicle-dynamics level.

\begin{assum}[Switched standstill hold]\label{ass:standstill}
Consider the dynamics~\eqref{eq:Sys} and set, with some abuse of notation, $\varphi_i(t) := \varphi_i(t,x_i(t),v_i(t),u_i(t))$ for $i=1,\ldots,N$ and $t\ge 0$. Each
follower $i=1,\ldots,N$ is equipped with a low-level standstill mechanism with two modes,
\(\mathcal S_i\in\{\mathrm{drive},\mathrm{hold}\}\), satisfying
\begin{equation}\label{eq:standstill}
\dot x_i(t)=v_i(t),
\qquad
m_i\dot v_i(t)=
\begin{cases}
\varphi_i(t), & \mathcal S_i(t)=\mathrm{drive},\\[2pt]
0, & \mathcal S_i(t)=\mathrm{hold}.
\end{cases}
\end{equation}
The mode is governed by the deterministic guards
\begin{equation}\label{eq:guards}
\begin{aligned}
\mathrm{drive}\to\mathrm{hold}:&\quad
v_i(t)=0\ \wedge\ \varphi_i(t)\le0,\\
\mathrm{hold}\to\mathrm{drive}:&\quad
\varphi_i(t)>0.
\end{aligned}
\end{equation}
The guards are understood as ideal continuous-time event detection at the
boundary \(v_i(t)=0\). Each guard depends only on the sign of $\varphi_i(t)$; for
this event detection to be well defined, we assume that 
the low-level standstill mechanism enforces a positive
dwell time: for each \(i=1,\ldots,N\), there exists \(\tau_{d,i}>0\) such that any two
consecutive standstill-mode switches of vehicle \(i\) are separated by at least
\(\tau_{d,i}\).

The initial mode is chosen consistently with the guards as
\[
\mathcal S_i(0)=
\begin{cases}
\mathrm{hold},&
v_i(0)=0\ \text{and}\ \varphi_i(0)\le0,\\
\mathrm{drive},&\text{otherwise}.
\end{cases}
\]
\end{assum}

\begin{rem}\label{rem:standstill_interpretation}
In Assumption~\ref{ass:standstill}, $\varphi_i(t)$ is the net longitudinal force before the standstill mechanism acts.
At zero velocity, the mechanism holds the vehicle while this force is nonpositive and releases it when the force becomes positive.
This gives an idealized description of brake-hold or hill-hold operation.
The low-level mechanism handles these switches, so the high-level controller does not need to calculate the net force.
The dwell-time assumption prevents arbitrarily rapid repeated switching.
\end{rem}

\begin{rem}\label{zeno}
The positive dwell time is an additional assumption on the low-level mechanism. It is consistent with practical brake-hold or
hill-hold systems, where hysteresis, filtering, sampling and
actuator response times prevent instantaneous repeated switching between hold
and drive. It therefore excludes Zeno behavior, meaning infinitely many switches in finite time. The solution can then be constructed by joining finitely many Carathéodory solution segments on each such interval.
\end{rem}

The standstill mechanism acts only at zero velocity.
Whenever $v_i(t)>0$, the vehicle follows the force-balance model~\eqref{eq:Sys}.

\subsection{Braking distance}

To guarantee collision avoidance under the input constraints, we first
introduce, in terms of the quantities from
Assumption~\ref{ass:bounds}, the worst-case normalized force contribution
\begin{align}\label{varpi}
    \varpi := \sin(\bar\theta)
    +\frac{\bar d}{\underline m g}\ge0.
\end{align}
We impose $\varpi<\underline c_d$. Indeed, under full braking,
$u_i=-c_{d,i}m_i g$, and therefore $g(c_{d,i}-\varpi)
\ge g(\underline c_d-\varpi)>0$ lower-bounds the available net deceleration of every vehicle. The rolling-friction contribution is deliberately omitted because
$F_{i,r}(v)=m_i g C_{i,r}s_\kappa(v)$ vanishes as $v\to0$ and hence
cannot be credited toward a uniform guaranteed deceleration. If
$\underline c_d\le\varpi$, the worst-case force contributions may
compensate for the guaranteed braking authority, and a uniform finite
stopping-distance bound cannot be ensured. A detailed derivation of the
corresponding worst-case braking distance for a single vehicle is
provided in~\cite{acc25}.

Building on this single-vehicle construction, we define the relative
braking-distance term
\begin{equation}\label{eq:db}
d_{b,i}(t)
=
\sigma_a\!\left(
\frac{v_i(t)^2-v_{i-1}(t)^2}
     {2g(\underline c_d-\varpi)}
\right),
\qquad t\ge0.
\end{equation}
where \(\sigma_a\in C^1(\mathbb R,\mathbb R_{\ge0}), a >0,\) is a one-sided approximation of \(r\mapsto\max\{0,r\}\)  satisfying
\begin{itemize}
    \item $\sigma_a(r)=0$ for all $r\le 0$,
    \item $0\le\sigma_a'(r)\le1$ for all $r\in\R$,
    \item $\lim_{a\to\infty} \sigma_a(r) = \max\{0,r\}$ for all $r\in\R$.
\end{itemize}
For example, the function $\sigma_a$ with \(\sigma_a(r)=r-\tfrac1a\bigl(1-e^{-ar}\bigr)\) for \(r>0\) and
\(\sigma_a(r)=0\) for \(r\le0\) satisfies these requirements.
The one-sided property \(\sigma_a(r)=0\) for \(r\le0\) is used in the hold-mode analysis. Notably, the above construction of $d_{b,i}$ avoids the use of the vehicle masses in the controller design, while still ensuring a conservative and implementable safety margin.

\subsection{High-level control design}

The overall decentralized (high-level) controller design for vehicle platoons~\eqref{eq:Sys} is given in the following, where $d_{b,i}(t)$ is defined in~\eqref{eq:db}:
\begin{equation}\label{eq:SSFCC}
\boxed{
\begin{aligned}
    e_i(t) &= x_i(t) - x_{i-1}(t)  + d_{b,i}(t) \\
    & \quad + \lambda v_i(t) + \frac{\mu v_i(t)^2}{\vmax-v_i(t)} + \psi^d_{f,i}\\[0.4em]
    \xi_i(t) &= \frac{e_i(t) - \frac{1}{2} \left( \psi^d_{f,i} + \psi^a_{i}(t) \right) }{\frac{1}{2} \left( \psi^d_{f,i} - \psi^a_{i}(t) \right) } \\[0.6em]
    \epsilon_i(t) &=  \ln{\left( \frac{1+\xi_i(t) }{1-\xi_i(t) } \right )},\quad \zeta_i(t) =  \frac{1}{1-\xi_i^2(t)} \geq 1 \\
    u_{d_i}(t)  &=  - k_i \zeta_i(t) \epsilon_i(t)\\[0.6em]
    u_{i}(t) &= \mathrm{sat}_{\underline{u}_i}^{\bar{u}_i}\!\big(u_{d_i}(t)\big) \\[0.6em]
   \dot \psi^a_{i}(t) &= - \ell_i^a \big(\psi^a_{i}(t)  + \psi^a_{f,i}\big)
    + s_{\gamma}(-\xi_i(t))
    \frac{u_{i}(t) - u_{d_i}(t)}{1-\xi_i(t) },\\
    \psi_i^a(0) &= -\psi^a_{f,i}
\end{aligned}
}
\end{equation}
with the controller design parameters
\begin{equation}\label{eq:FC-param}
\boxed{
\begin{aligned}
& \psi^a_{f,i}>0,\quad \ell_i^a>0,\ \psi^d_{f,i}>0,\\
& k_i >0,\quad \lambda>0,\quad \mu>0,\quad \vmax>0,\quad \gamma>0, \\
& s_\gamma:\R\to \R_{\ge 0},\ \chi\mapsto  \begin{cases} \gamma, & \chi \geq 0 \\ 0, & \chi< 0 \end{cases}
\end{aligned}
}
\end{equation}
for $i=1,\ldots,N$ and the saturation function
\[
    \mathrm{sat}_{\underline{u}_i}^{\bar{u}_i}:\R\to\R,\ u\mapsto  \begin{cases} \bar{u}_i, & u> \bar{u}_i \\ u, & \underline{u}_i \le u \le \bar{u}_i\\ \underline{u}_i, & u < \underline{u}_i \end{cases}
\]
associated with the control bounds~\eqref{eq:control_bounds}.

We note that the performance function $\psi_i^a$, which guarantees collision avoidance, is dynamically generated based on the difference between the control signal $u_{d_i}$ and the control input $u_i$, which is a saturated version of the former and satisfies the input constraints. Whenever saturation is active, a negative contribution is added to $\dot \psi_i^a(t)$, thus inducing a widening of the error margin. If saturation is inactive, $\psi_i^a$ recovers to its desired shape (prescribed by the parameters $\ell_i^a$ and $\psi_{f,i}^a$) exponentially fast. The relaxation of the performance function is the crucial technique which allows to simultaneously achieve collision avoidance and satisfy the input constraints. Earlier works on prescribed performance control and funnel control under input constraints already implemented similar relaxation techniques~\cite{Berg24,Berg26,TrakBech23,TrakBech24}.

Note that the initial value of $\psi_i^a$ is chosen as its nominal value, $\psi_i^a(0) = -\psi^a_{f,i}$; by Remark~\ref{rem:dormant}
below, $\psi_i^a$ then coincides with a constant funnel as long as the
propulsion constraint is inactive. The more general choice
$\psi_i^a(0)<-\psi^{a}_{f,i}$ of \cite{Berg24} is possible at the cost of
an additional upper bound on $\ell^{a}_i$, cf.\ Remark~\ref{rem:general-psi0}.

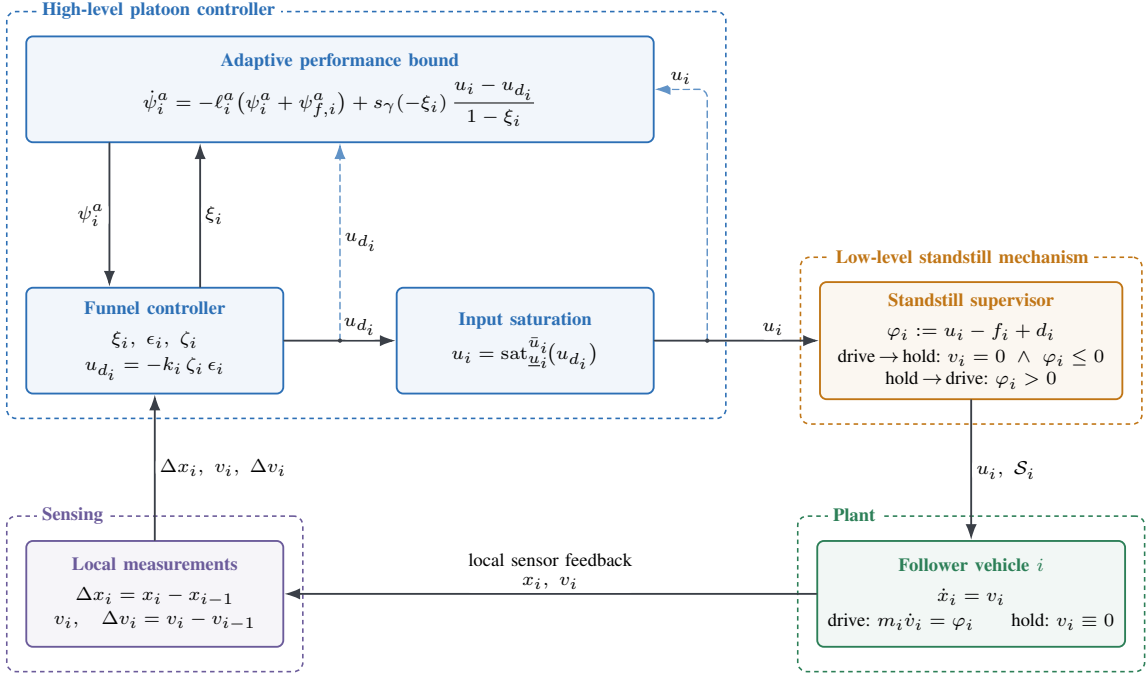
\begin{figure*}[thpb]
\centering
\usetikzlibrary{arrows.meta,positioning,fit,calc,backgrounds}

\definecolor{cHigh}{HTML}{2C6FB5}   
\definecolor{cLow}{HTML}{C07515}    
\definecolor{cPlant}{HTML}{2C7F57}  
\definecolor{cSens}{HTML}{6B5B9A}   
\definecolor{cWire}{HTML}{3A3F47}   

\newcommand{\blktitle}[2]{\textcolor{#1}{\bfseries #2}}

\begin{tikzpicture}[
  font=\scriptsize,
  >={Latex[length=2.1mm,width=1.5mm]},
  line join=round, line cap=round,
  blk/.style={draw=#1, line width=0.7pt, fill=#1!6,
              rounded corners=2.5pt, align=center, inner sep=4pt,
              minimum width=34mm, minimum height=14mm},
  blk/.default=cWire,
  wire/.style={draw=cWire, line width=0.7pt, rounded corners=3pt, ->},
  tap/.style ={draw=cHigh!75, line width=0.55pt, rounded corners=3pt, ->,
               dash pattern=on 2.4pt off 1.6pt},
  lbl/.style ={inner sep=1.6pt, fill=white, font=\scriptsize},
  dot/.style ={circle, fill=cWire, inner sep=0pt, minimum size=1.6pt},
  grp/.style={draw=#1, dash pattern=on 1.7pt off 1.5pt, line width=0.6pt,
              rounded corners=3.5pt, inner xsep=7pt, inner ysep=9pt},
  grp/.default=cWire,
  gttl/.style={font=\scriptsize\bfseries, text=#1, fill=white, inner sep=2pt},
  gttl/.default=cWire
]

\node[blk=cHigh] (ctrl) at (0,0) {%
  \blktitle{cHigh}{Funnel controller}\\[1.2mm]
  $\xi_i,\ \epsilon_i,\ \zeta_i$\\[0.6mm]
  $u_{d_i}=-k_i\,\zeta_i\,\epsilon_i$};

\node[blk=cHigh] (sat) at (4.9,0) {%
  \blktitle{cHigh}{Input saturation}\\[1.2mm]
  $u_i=\operatorname{sat}_{\underline{u}_i}^{\bar u_i}\!\bigl(u_{d_i}\bigr)$};

\node[blk=cLow, minimum width=40mm] (sup) at (10.8,0) {%
  \blktitle{cLow}{Standstill supervisor}\\[1.2mm]
  $\varphi_i:=u_i-f_i+d_i$\\[0.6mm]
  drive\,$\to$\,hold:\ $v_i=0\ \wedge\ \varphi_i\le0$\\[0.2mm]
  hold\,$\to$\,drive:\ $\varphi_i>0$};

\node[blk=cPlant, minimum width=40mm] (veh) at (10.8,-3.35) {%
  \blktitle{cPlant}{Follower vehicle $i$}\\[1.2mm]
  $\dot x_i=v_i$\\[0.4mm]
  drive:\ $m_i\dot v_i=\varphi_i$\qquad hold:\ $v_i\equiv0$};

\node[blk=cSens] (meas) at (0,-3.35) {%
  \blktitle{cSens}{Local measurements}\\[1.2mm]
  $\Delta x_i=x_i-x_{i-1}$\\[0.4mm]
  $v_i,\quad \Delta v_i=v_i-v_{i-1}$};

\node[blk=cHigh, anchor=south west,
      minimum width={83mm}] (psi) at ($(ctrl.north west)+(0,19mm)$) {%
  \blktitle{cHigh}{Adaptive performance bound}\\[1.2mm]
  $\dot\psi_i^{a}=-\ell_i^{a}\bigl(\psi_i^{a}+\psi_{f,i}^{a}\bigr)
   +s_\gamma(-\xi_i)\,\dfrac{u_i-u_{d_i}}{1-\xi_i}$};

\draw[wire] (ctrl.east) -- node[lbl,above=0pt,pos=0.68]{$u_{d_i}$} (sat.west);
\draw[wire] (sat.east)  -- node[lbl,above=0pt,pos=0.72]{$u_i$}    (sup.west);
\draw[wire] (sup.south) -- node[lbl,right=0pt]{$u_i,\ \mathcal{S}_i$} (veh.north);
\draw[wire] (veh.west)  -- node[lbl,above=0pt,align=center]
                           {local sensor feedback\\[-0.4mm]$x_i,\ v_i$} (meas.east);
\draw[wire] (meas.north)-- node[lbl,right=0pt]{$\Delta x_i,\ v_i,\ \Delta v_i$} (ctrl.south);

\coordinate (xPsi) at ($(ctrl.north)+(-6mm,0)$);   
\coordinate (xXi)  at ($(ctrl.north)+( 6mm,0)$);   
\draw[wire] (psi.south -| xPsi) -- node[lbl,left=0pt] {$\psi_i^{a}$} (xPsi);
\draw[wire] (xXi) -- node[lbl,right=0pt]{$\xi_i$} (psi.south -| xXi);

\coordinate (tud) at ($(ctrl.east)!0.5!(sat.west)$);
\coordinate (tui) at ($(sat.east)!0.32!(sup.west)$);
\draw[tap] (tud) -- node[lbl,right=0pt]{$u_{d_i}$} (psi.south -| tud);
\draw[tap] (tui) -- (tui |- psi.east) -- node[lbl,above=0pt]{$u_i$} (psi.east);
\node[dot] at (tud) {};
\node[dot] at (tui) {};

\begin{scope}[on background layer]
  \node[grp=cHigh,  fit=(ctrl)(sat)(psi)(tui)] (gHigh) {};  
  \node[grp=cLow,   fit=(sup)]                 (gLow)  {};
  \node[grp=cPlant, fit=(veh)]                 (gVeh)  {};
  \node[grp=cSens,  fit=(meas)]                (gSens) {};
\end{scope}

\node[gttl=cHigh,  anchor=west] at ($(gHigh.north west)+(4mm,0)$) {High-level platoon controller};
\node[gttl=cLow,   anchor=west] at ($(gLow.north west)+(4mm,0)$)  {Low-level standstill mechanism};
\node[gttl=cPlant, anchor=west] at ($(gVeh.north west)+(4mm,0)$)  {Plant};
\node[gttl=cSens,  anchor=west] at ($(gSens.north west)+(4mm,0)$) {Sensing};

\end{tikzpicture}
 \caption{Signal-flow structure of the high-level platoon controller
    with input saturation, auxiliary compensation dynamics, low-level
    standstill mechanism, and follower-vehicle dynamics. }
    \label{fig:standstill-controller}
\end{figure*}

The overall architecture of the proposed controller~\eqref{eq:SSFCC} is illustrated in Fig.~\ref{fig:standstill-controller}. The measured or estimated signals \(d_i\), $v_i$, and \(v_{i-1}\) are used by the high-level platoon controller to compute the desired input \(u_{d_i}\).  After saturation, the admissible input \(u_i\) is passed to the (low-level) standstill mechanism, which determines whether the follower vehicle evolves in drive mode or hold mode. 

\section{Stability analysis}\label{stability}
Before stating the main result, we introduce some assumptions on the system and design parameters as well as the initial conditions, necessary for the stability proof. 

\begin{assum}\label{Ass:design-param}
We assume that the controller design parameters~\eqref{eq:FC-param}, the actuator constraints~\eqref{eq:control_bounds} and the initial conditions~\eqref{eq:IC} satisfy the following conditions, which can be checked successively:
\begin{enumerate}[label=(P\arabic*),leftmargin=*]
\item\label{P1}\emph{(Actuator authority)}
  $\varpi<\underline c_d$ and $\bar c_a+\varpi\le\underline c_d$.
\item\label{P2}\emph{(Design speed)}
Let $\bar v_0, \bar a_0$ be as in Assumption~\ref{ass:leader}. We have $v_i(0)\ge0$ for $i=1,\ldots,N$ and, for
\begin{align*}
    \underline v &:=\max\Bigl\{\bar v_0,\;
     \sup_{N\in\mathbb N} \max_{1\le i\le N}v_i(0)\Bigr\},\\
     v_c&:= \sqrt{\frac{\underline m\,g\bigl(\underline c_{a}-\sin\bar\theta-\bar C_r\bigr)-\bar d}{\bar c_\rho}}
\end{align*}
 we have that $\underline v<v_c$. Then choose $\vmax\in(\underline v,\,v_c)$, which determines the constants
  \begin{align*}
    \underline\alpha&:=g\bigl(\underline c_{a}-\sin\bar\theta-\bar C_r\bigr)
            -\frac{\bar d+\bar c_\rho\vmax^{2}}{\underline m}>0,\\
    \bar a&:=
g(\bar c_d+\bar C_r+\sin\bar\theta)
+\frac{\bar d}{\underline m}
+\frac{\bar c_\rho}{\underline m}\vmax^2,\\ 
    \beta&:=\vmax\Bigl(1+\frac{\max\{\bar a,\bar a_0\}}{g(\underline c_d-\varpi)}\Bigr).
  \end{align*}
\item\label{P3}\emph{(Time headway)}
  $\lambda>\beta/\underline\alpha$.
\item\label{P4}\emph{(Barrier gain)}
  For $\Gamma^{0}_i:=x_{i-1}(0)-x_i(0)-d_{b,i}(0)$ we have that $\Gamma^{0}_i>\lambda v_i(0)$ for all
  $i=1,\ldots, N$. Then choose 
  \[
    \mu<    \bar\mu:= \inf_{N\in\mathbb N} \min_{\substack{1\le i\le N\\ v_i(0)>0}}
      \frac{\bigl(\Gamma^{0}_i-\lambda v_i(0)\bigr)\bigl(\vmax-v_i(0)\bigr)}
           {v_i(0)^{2}}.
    \]
\item\label{P5}\emph{(Funnel boundaries and initial conditions)}
For $i=1,\ldots,N$, let $r_i^a>0$ be the unique solution of $\mathcal T(r_i^a)=\frac{c_{a,i}m_i g}{k_i}$, \(\mathcal T:(0,\infty)\to(0,\infty),\ r\mapsto r\cosh^2\!\left(\frac r2\right)\) and set
\[
\phi_i^a :=
1+\tanh\!\left(\frac{r_i^a}{2}\right)\in(1,2)
\] 
as well as the admissible initial funnel excursion
\begin{align*}
  \bar\xi^{\,0}_i&:=\tanh\Bigl(\tfrac{\bar\epsilon^{\,0}_i}{2}\Bigr)\in(0,1),\\
    \bar\epsilon^{\,0}_i&:=\mathcal T^{-1}
      \Bigl(\frac{c_{a,i}m_ig}{k_i}
            +\frac{2(\lambda\underline\alpha-\beta)\,\phi^{a}_i}{(2-\phi^{a}_i)\,\gamma k_i}\Bigr)
      >r^{a}_i .
\end{align*}
Set $G_i:= \Gamma^{0}_i-\lambda v_i(0) - \tfrac{\mu v_i(0)^2}{\vmax-v_i(0)}$ and assume that for
  \[
    \underline\psi^{\,d}_i:=\max\Bigl\{0,\
      \frac{2G_i}{1+\bar\xi^{\,0}_i}-\psi^a_{f,i}\Bigr\},\quad
    \bar\psi^{\,d}_i:=\frac{2G_i}{1-\bar\xi^{\,0}_i}-\psi^a_{f,i} 
  \]
  we have $\underline\psi^{\,d}_i<\bar\psi^{\,d}_i$. Then choose
  $\psi^{d}_{f,i}\in\Bigl(\underline\psi^{\,d}_i,\ \bar\psi^{\,d}_i\Bigr)$.
\item\label{P6}\emph{(Uniformity in $N$)}
The vehicle-dependent controller parameters are uniformly bounded with respect to $N$, namely,
\begin{align*}
0&<
\inf_{N\in\mathbb N}\min_{1\le i\le N} (\psi^{a}_{f,i}+\psi^{d}_{f,i} + k_i +\ell_i^a)\\ 
&\le
\sup_{N\in\mathbb N}\max_{1\le i\le N}(\psi^{a}_{f,i}+\psi^{d}_{f,i} + k_i +\ell_i^a) <\infty
\end{align*}
\end{enumerate}
\end{assum}

\begin{rem}[Structure of the conditions]\label{rem:structure}
Assumption~\ref{Ass:design-param} provides a step-by-step procedure for choosing the controller parameters.
Each step uses quantities fixed in the preceding steps and specifies an explicit admissible range. The time-headway condition~\ref{P3}, $\lambda>\beta/\underline\alpha$, can be conservative because it uses worst-case velocity and acceleration bounds and the smallest guaranteed propulsion margin.
It is sufficient for the theoretical guarantees, but smaller headways may also work.
The simulations in Section~\ref{Sec:Sim} show safe operation and bounded velocity deviations with a much smaller headway, despite large differences in vehicle mass and time-varying disturbances.
\end{rem}
\begin{rem}[Velocity bound]\label{rem:vmax}
The barrier term $\mu v_i^2/(\vmax-v_i)$ extends the linear spacing policy of~\cite{BergBess26} and grows without bound as $v_i$ approaches $\vmax$.
Keeping the error below the upper funnel boundary therefore prevents the velocity from reaching $\vmax$, provided $v_i(0)<\vmax$.
The funnel thus enforces a prescribed speed limit.
Condition~\ref{P2} ensures that a uniform propulsion margin $\underline\alpha>0$ is feasible at that speed.
\end{rem}
\begin{rem}[Effective time headway]\label{rem:time_headway}
The speed-dependent spacing terms in~\eqref{eq:SSFCC} correspond to an effective time headway $\lambda+\tfrac{\mu v_i}{\vmax-v_i}.$ This is close to $\lambda$ at low speeds and increases as $v_i$ approaches $\vmax$.
It reaches $2\lambda$ at $v_i=\lambda\vmax/(\lambda+\mu)$.
For any fixed $v_i<\vmax$, the linear spacing policy $\lambda v_i$ is recovered as $\mu\to0$.
\end{rem}
\begin{rem}[Initial funnel width]\label{rem:P5}
By~\eqref{eq:SSFCC} and the definitions in~\ref{P4}--\ref{P5}, the initial normalized error satisfies
\begin{equation}\label{eq:xi0}
  \xi_i(0)=1-\frac{2G_i}{\psi^{d}_{f,i}+\psi^{a}_{f,i}}
\end{equation}
with $G_i$ as defined in \ref{P5}.
For a fixed $G_i$, the initial funnel width determines $\xi_i(0)$.
The interval in~\ref{P5} ensures $|\xi_i(0)|<\bar\xi^{\,0}_i<1$, so the initial error lies strictly inside the funnel.
A simpler sufficient choice is
\[
  \psi^{d}_{f,i}+\psi^{a}_{f,i}
  \in\left[
    \frac{2G_i}{\phi^{a}_i},
    \frac{2G_i}{2-\phi^{a}_i}
  \right].
\]
This gives $|\epsilon_i(0)|\le r^{a}_i$ and avoids initial propulsion saturation.
\end{rem}
\begin{rem}[Adaptation and propulsion saturation]\label{rem:dormant}
With $\psi^{a}_i(0)=-\psi^{a}_{f,i}$, the funnel boundaries remain constant until the requested propulsion exceeds the available force.
Only then does the saturation term in~\eqref{eq:SSFCC} relax the lower boundary.
Once propulsion saturation ends, this boundary returns exponentially toward its nominal value, as long as propulsion remains unsaturated.
Braking saturation does not trigger funnel relaxation.
\end{rem}
\begin{rem}[General initial funnel width]\label{rem:general-psi0}
A wider initial funnel can be chosen by setting $\psi^{a}_i(0)<-\psi^{a}_{f,i}$.
Condition~\ref{P5} must then be supplemented by
\[
  \ell^{a}_i<
  \frac{2(\lambda\underline\alpha-\beta)}
       {(2-\phi^{a}_i)
        \bigl(|\psi^{a}_i(0)|-\psi^{a}_{f,i}\bigr)}.
\]
Also, $\psi^{a}_{f,i}$ must be replaced by $|\psi^{a}_i(0)|$ in the definitions of $\underline\psi^{\,d}_i$ and $\bar\psi^{\,d}_i$, and in~\eqref{eq:xi0}.
The nominal initialization $\psi^{a}_i(0)=-\psi^{a}_{f,i}$ avoids this additional restriction on the recovery rate $\ell^{a}_i$.
\end{rem}
\begin{rem}[On the conservatism of $\underline\alpha$]\label{rem:alpha-sharp}
The bound in~\ref{P2} combines the smallest propulsion coefficient with the smallest mass, even when they belong to different vehicles.
This can underestimate the available propulsion margin and increase the required time headway.
A less conservative alternative is to choose $\vmax>\underline v$ and a uniform $\underline\alpha>0$ satisfying
\[
  g(c_{a,i}-\sin\bar\theta-\bar C_r)
  -\frac{\bar d+\bar c_\rho\vmax^2}{m_i}
  \ge\underline\alpha
\]
for every vehicle and every platoon length.
This replaces the speed-feasibility test and the definition of $\underline\alpha$ in~\ref{P2}.
The other requirements and the definitions of $\bar a$ and $\beta$ remain unchanged.
The stability proof still applies because it uses only the positive uniform lower bound $\underline\alpha$.
\end{rem}

Next, we introduce the notion of a solution to the closed-loop system, consisting of the platoon dynamics~\eqref{eq:Sys} under input constraints~\eqref{eq:actuator_constraints} and the low- and high-level controllers~\eqref{eq:standstill} and~\eqref{eq:SSFCC}. We note that, in view of the standstill mechanisms~\eqref{eq:standstill} with two switching modes, the resulting solution will be a switching solution. Therefore, for a platoon of length~$N$, $(x,v,\psi,\cS) = (x_1,\ldots,x_N,v_1,\ldots,v_N,\psi_1^a,\ldots,\psi_N^a,\cS_1,\ldots,\cS_N):[0,\omega)\to\R^{4N}$, $\omega\in(0,\infty]$, is called a solution of~\eqref{eq:Sys},~\eqref{eq:actuator_constraints},~\eqref{eq:SSFCC}, if the initial conditions in~\eqref{eq:IC},~\eqref{eq:SSFCC} hold and $(x,v,\psi)$ is locally absolutely continuous and satisfies the differential equations in~\eqref{eq:Sys} and~\eqref{eq:SSFCC} with $u_1,\ldots,u_N$ as in~\eqref{eq:SSFCC} and modes $\cS_1,\ldots,\cS_N$ (with only finitely many jumps in any compact interval) determined by~\eqref{eq:guards} for almost all $t\in [0,\omega)$; $(x,v,\psi,\cS)$ is called maximal, if it has
no right extension that is also a solution. 

To state the velocity bound, define the speed-dependent spacing function
\[
  h:[0,\vmax)\to[0,\infty),\qquad
  h(v):=\lambda v+\frac{\mu v^2}{\vmax-v},
\]
and the guaranteed braking deceleration
$\delta:=g(\underline c_d-\varpi)>0$.
The function $h$ is strictly increasing and has an inverse
$h^{-1}:[0,\infty)\to[0,\vmax)$.

For a solution on $[0,\omega)$, $\omega\in(0,\infty]$ define
\begin{align*}
  n_i(t)&:=\min\Big\{
    \min_{s\in[0,t]}v_{i-1}(s),\,v_i(0)
  \Big\},\\
  N_i(t)&:=\max\Big\{
    \max_{s\in[0,t]}v_{i-1}(s),\,v_i(0)
  \Big\},\\
  W_i&:=\psi^d_{f,i}
    +\sup_{0\le s<\omega}\bigl(-\psi^a_i(s)\bigr).
\end{align*}
The interval $[n_i(t),N_i(t)]$ contains the predecessor's
velocity history up to time $t$ and the follower's initial velocity.
The quantity $W_i$ is the supremum of the funnel width along
the solution.

We are now in the position to state the main result of this paper.

\begin{thm}\label{thm:main}
Consider a platoon of~$N$ vehicles with dynamics~\eqref{eq:Sys} and initial conditions~\eqref{eq:IC}, subject to input constraints~\eqref{eq:actuator_constraints} and under control~\eqref{eq:standstill},~\eqref{eq:SSFCC}, such that Assumptions~\ref{ass:leader},~\ref{ass:bounds},~\ref{ass:standstill} and~\ref{Ass:design-param} hold. Then the closed-loop system~\eqref{eq:Sys},~\eqref{eq:standstill},~\eqref{eq:SSFCC} has a solution, and every solution can be extended to a maximal solution $(x_1,\ldots,x_N,v_1,\ldots,v_N,\psi_1^a,\ldots,\psi_N^a,\cS_1,\ldots,\cS_N):[0,\omega)\to\R^{4N}$, $\omega\in(0,\infty]$, which has the properties:
\begin{enumerate}[label=(\roman*),leftmargin=*]
  \item global existence: $\omega=\infty$;
  \item collision avoidance: $x_i(t)-x_{i-1}(t) < 0$ for all $t\ge 0$;
  \item boundedness and speed limitation:
  $v_i(t)\in[0,\vmax)$ for all $t\ge0$, and $\epsilon_i$, $u_{d_i}$  and $\psi^{a}_i$ are bounded for all $i=1,\ldots,N$, independent of~$i$ and~$N$;
  \item link-wise velocity string stability: there is a constant $W>0$, independent of $i$ and $N$, such that $W_i\le W$.
For every $i=1,\ldots,N$ and $t\ge0$ we have:
\begin{equation}\label{eq:linkwise-thm}
\begin{aligned}
  &h^{-1}\Big(\max\big\{h(n_i(t))-W_i,0\big\}\Big)\\
  &\quad\le v_i(t)<
  h^{-1}\Big(h(N_i(t))+W_i+\frac{N_i(t)^2}{2\delta}\Big).
\end{aligned}
\end{equation}
Consequently,~\eqref{eq:string-stab} holds with
\[
  \Theta=\frac{W+\vmax^2/(2\delta)}{\lambda}.
\]
      
\end{enumerate}
\end{thm}
\begin{proof}
The proof first establishes local existence and shows that the standstill mechanism keeps the velocities nonnegative.
It then derives uniform bounds on the controller signals and shows that the errors remain strictly inside the funnel.
Together with the speed barrier, these estimates establish collision avoidance and allow every solution to be continued for all time.
Finally, the relation between spacing and velocity yields the link-wise bound~\eqref{eq:linkwise-thm}.
The complete proof is given in Appendix~\ref{app:proof-main}.
\end{proof}

\begin{rem}[Acceleration bound]\label{rem:collapse}
Under~\ref{P1}, the deceleration bound derived in Step~3 of the proof is no smaller than the acceleration bound.
Thus, the expression for $\bar a$ in~\ref{P2} bounds both acceleration and deceleration.
The second condition in~\ref{P1} requires the smallest normalized braking limit in the platoon to be at least the largest normalized propulsion limit plus the worst-case load contribution $\varpi$.
\end{rem}

\begin{rem}[Link-wise velocity string stability]\label{rem:pvss}
Assertion~(iv) of Theorem~\ref{thm:main} states that each follower tracks the velocity range of its predecessor
with gain one, up to an offset which, measured in units of the spacing function~$h$,
consists of the tube width~$W_i$ and, on the upper side, of the braking distance
$N_i(t)^2/(2\delta)$ that becomes available when the predecessor accelerates. The
lower bound excludes, in particular, that a follower comes to rest while its
predecessor keeps driving. Applying~\eqref{eq:linkwise-thm} successively along the
platoon yields, with
$\underline v_0:=\inf_{t\ge0}v_0(t)$,
$\underline v_0^*:=\min\{\underline v_0,\min_{1\le j\le i}v_j(0)\}$,
$\bar v_0^*:=\max\{\bar v_0,\max_{1\le j\le i}v_j(0)\}$
and $c:=W+\vmax^2/(2\delta)$, that
\begin{align*}
  |v_i(t)-v_0(t)|\le\max\Big\{\,&
  h^{-1}\big(h(\bar v_0^*)+ic\big)-v_0(t),\\
  &\bar v_0-h^{-1}\big(
    \max\{h(\underline v_0^*)-iW,0\}
  \big)\Big\}
\end{align*}
for all $t\ge 0$, and the right-hand side is bounded both by
$(\bar v_0-\underline v_0)+I_i+ic/\lambda$, with $I_i$ as in Section~\ref{Ssec:ContrObj} and by
the speed limit: because of the barrier term in~$h$, the bound on $\sup_t v_i(t)$
approaches~$\vmax$ from below as $i$ increases and never exceeds it. The estimate is most informative for vehicles near the leader and when the leader's speed range $\bar v_0-\underline v_0$ is small compared with $\vmax$.
If the leader stops after travelling close to $\vmax$, the estimate may approach the speed limit.
This reflects the fact that followers can still be moving after the leader has stopped. The dependence of the offset on $i$ remains because the argument uses only the spacing constraint~\eqref{eq:tube}.
This constraint allows excursions toward the far ends of the tubes to occur at different times and accumulate toward the rear of the platoon.
Even with a leader travelling at constant speed, such excursions can produce velocity deviations of order $\sqrt{N}\,W/\lambda$ over a range of platoon lengths.
This growth is ultimately limited by the prescribed speed bound. A bound that is uniform
in~$N$ would have to exploit the closed-loop dynamics inside the tube and is beyond
the scope of the present paper.
\end{rem}

\section{Simulation}\label{Sec:Sim}

We consider two heterogeneous platoons undergoing a stop-and-go
maneuver. Case~1 satisfies \ref{P1}-\ref{P6} and includes brief
actuator saturation. Case~2 uses a smaller time headway, outside
the sufficient conditions. We also examine longer platoons and
compare the controller with~\cite{BergBess26} under identical
actuator limits.

\emph{Common setup.}
Both cases use $N=20$ followers, with
$\tfrac12\rho_i C_{i,d}A_i=0.4992~\si{\kilo\gram\per\metre}$,
$C_{i,r}=0.01$, and $\kappa=10$.
The disturbances are $d_i(t)=\bar d\sin(0.4t+\phi_i)$,
with $\phi_i=2\pi(i-1)/20$.
We set $\psi_{f,i}^d=0.5~\si{\metre}$,
$\psi_{f,i}^a=0.1~\si{\metre}$,
$\psi_i^a(0)=-\psi_{f,i}^a$, $k_i=5m_i$, and $\ell_i^a=40$.
For $\sigma_a$, we use the exponential choice with $a=50$.
Initially, the velocities and inter-vehicle distances are uniform.
Shading in the trajectory-figure insets marks actuator saturation.

\emph{Case~1: Saturation within the sufficient conditions.}
\begin{figure}[h]
  \centering
  \includegraphics[width=\columnwidth]{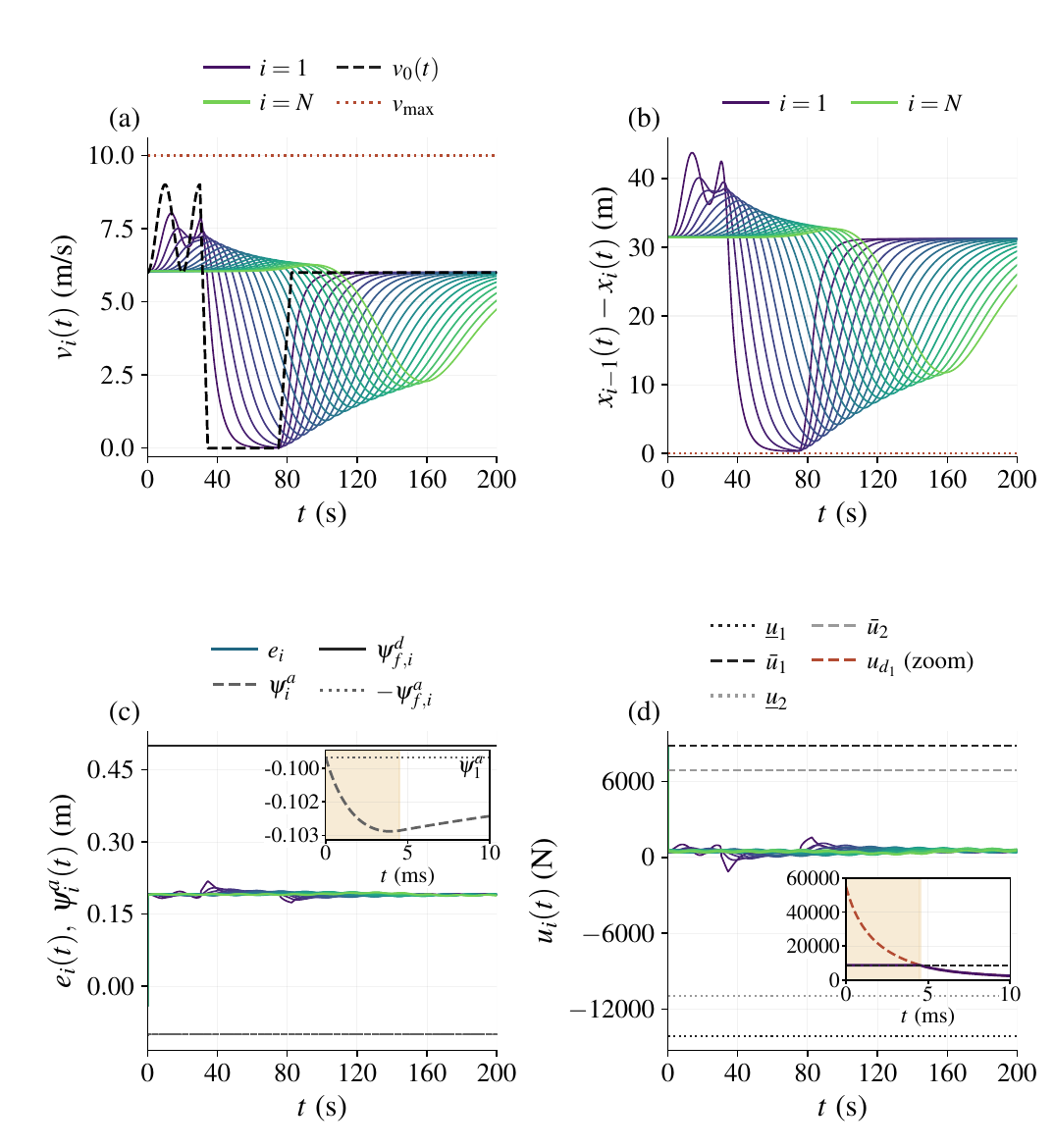}
  \caption{Case~1: velocities, gaps, funnel errors, and applied forces.
  Insets: first $10~\mathrm{ms}$.}
  \label{fig:case1_trajectories}
\end{figure}

\begin{figure}[h]
  \centering
  \includegraphics[width=\columnwidth]{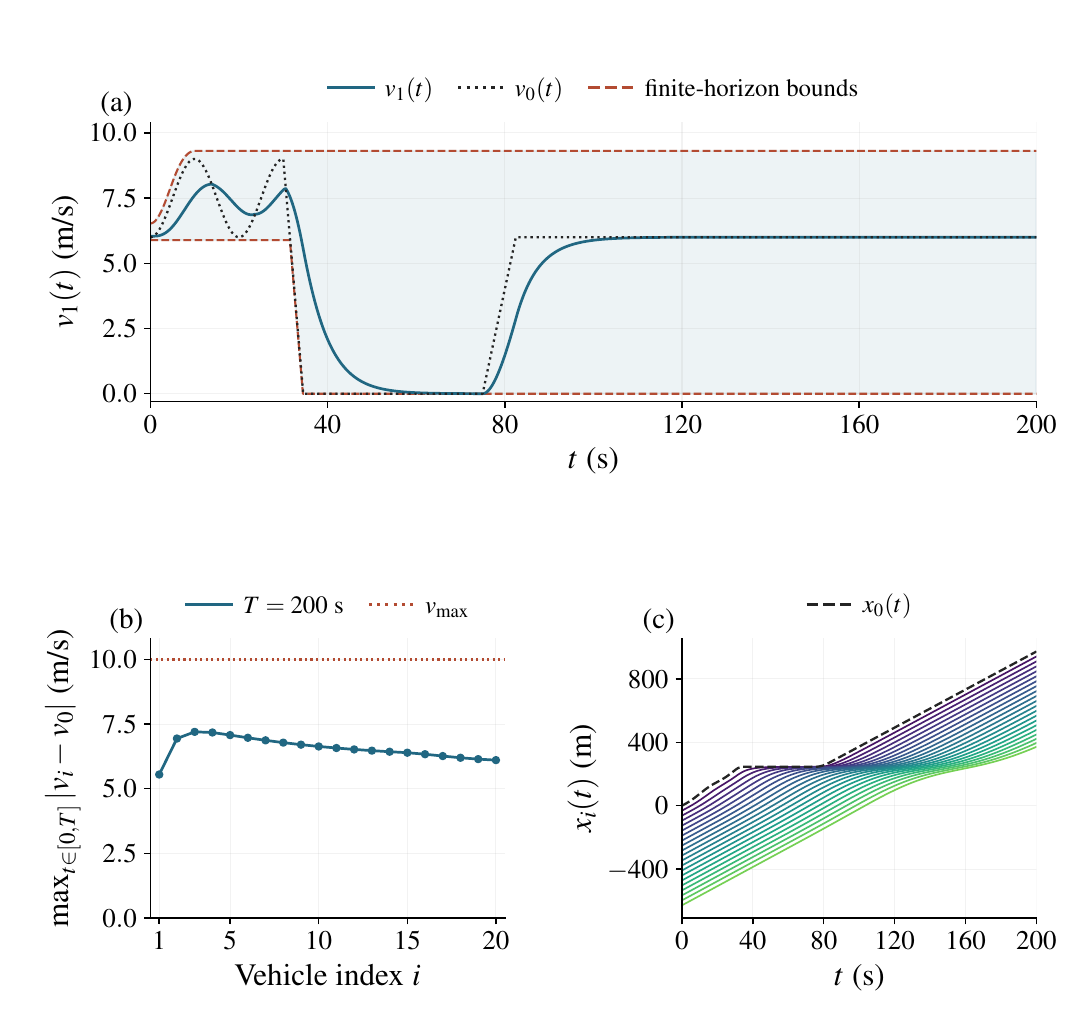}
  \caption{Case~1: velocity envelopes, peak velocity deviations, and positions.}
  \label{fig:case1_diagnostics}
\end{figure}
The masses are $m_i=1600+(-1)^{i+1}200~\si{\kilo\gram}$,
so the first follower is heavy. We use $c_d=0.8$,
$c_{a,i}=0.5$, $\theta=0.02~\si{\radian}$, and
$\bar d=50~\si{\newton}$.
We choose $\vmax=10~\si{\metre\per\second}$,
$\mu=0.1$, $\lambda=5~\si{\second}$, and $\gamma=10^{-4}$.
The initial velocity is $6~\si{\metre\per\second}$ and the
spacing is $31.44~\si{\metre}$, giving $\xi_i(0)=-0.8$.
For \ref{P1}, $\varpi\approx0.02364$ and
$0.5+\varpi<0.8$. Condition~\ref{P2} holds because
$\bar v_0=9<\vmax=10<v_c\approx113.27$, with
$\underline\alpha\approx4.539$.
Moreover, \ref{P3}--\ref{P5} give, respectively,
\begin{align*}
 \beta/\underline\alpha&\approx4.579<\lambda,\\
 \mu&<\bar\mu=0.16,\\
 |\xi_i(0)|&<\min_j\bar\xi_j^0\approx0.8678.
\end{align*}
Condition~\ref{P6} holds because the same two vehicle types
and their positive controller parameters are repeated as $N$ grows.

For the first $30~\si{\second}$, the leader follows
$v_0(t)=6+1.5(1-\cos(\pi t/10))$.
It then brakes at $2~\si{\metre\per\second\squared}$,
stops at $34.5~\si{\second}$, and waits until $75~\si{\second}$.
It restarts at $0.8~\si{\metre\per\second\squared}$ to
$6~\si{\metre\per\second}$.
Over $T=200~\si{\second}$, the gaps remain positive,
the velocities stay below $\vmax$, and the errors remain inside
their funnels (Fig.~\ref{fig:case1_trajectories}).
Initial propulsion saturation lasts approximately
$4.5$--$4.8~\si{\milli\second}$ across the platoon.
For the first follower, $u_1=8829~\si{\newton}$ during saturation
and $\psi_1^a$ decreases to approximately $-0.1033~\si{\metre}$.
The insets show this small relaxation over the first
$10~\si{\milli\second}$.
No follower enters hold mode during this run.
Figure~\ref{fig:case1_diagnostics} also shows finite-horizon
velocity envelopes inspired by assertion~(iv). These envelopes use
the observed tube width
$W_{1,T}^{\mathrm{obs}}=\psi_{f,1}^d+\max_{t_k\le T}(-\psi_1^a(t_k))$
at the sampling times $t_k$. They visualize the structure of
the link-wise estimate, but are not independent a priori certificates.

\emph{Case~2: Smaller headway and greater mass heterogeneity.}
\begin{figure}[t]
  \centering
  \includegraphics[width=\columnwidth]{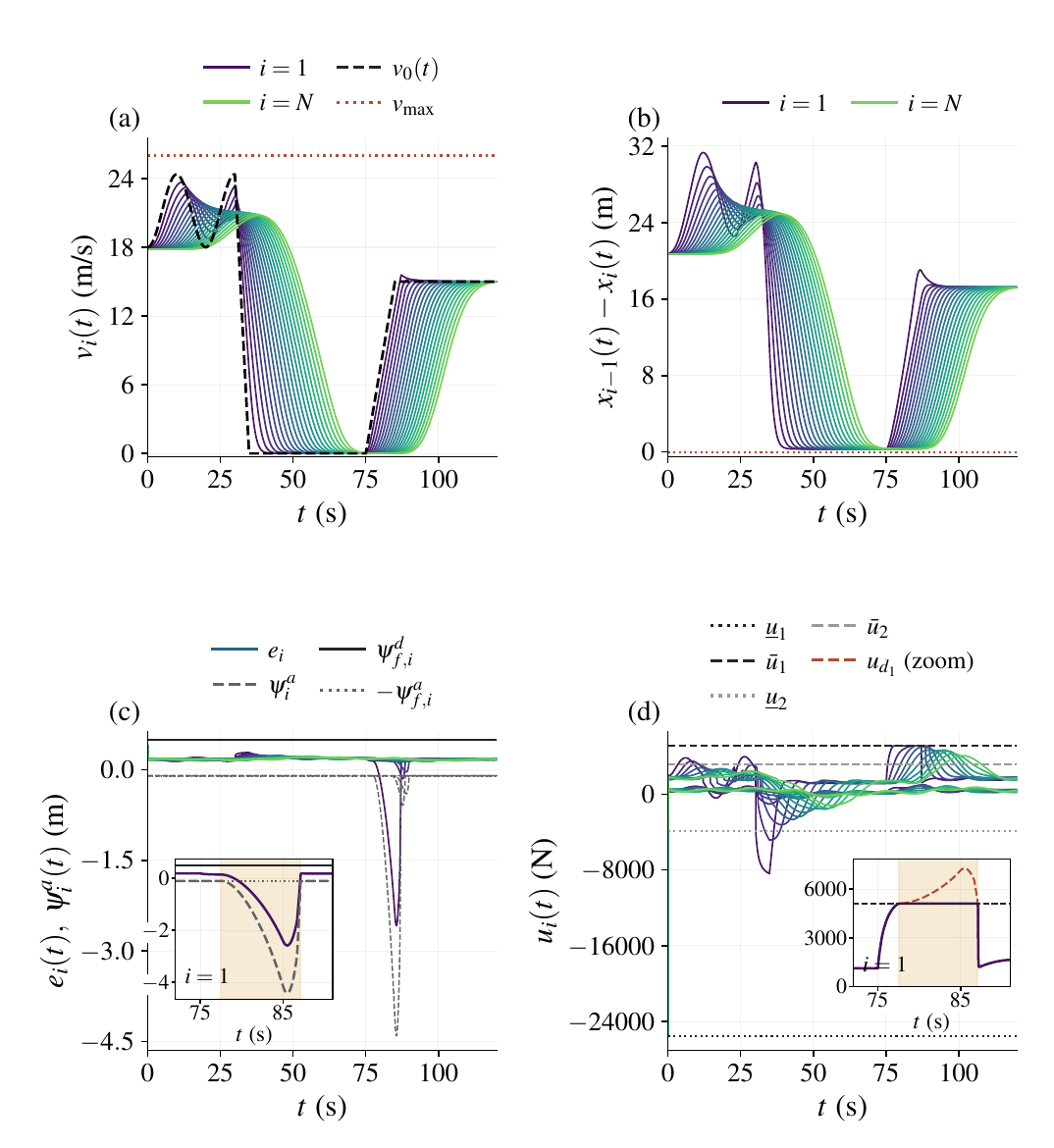}
  \caption{Case~2: velocities, gaps, funnel errors, and applied forces.
  Insets: restart.}
  \label{fig:case2_trajectories}
\end{figure}

\begin{figure}[t]
  \centering
  \includegraphics[width=\columnwidth]{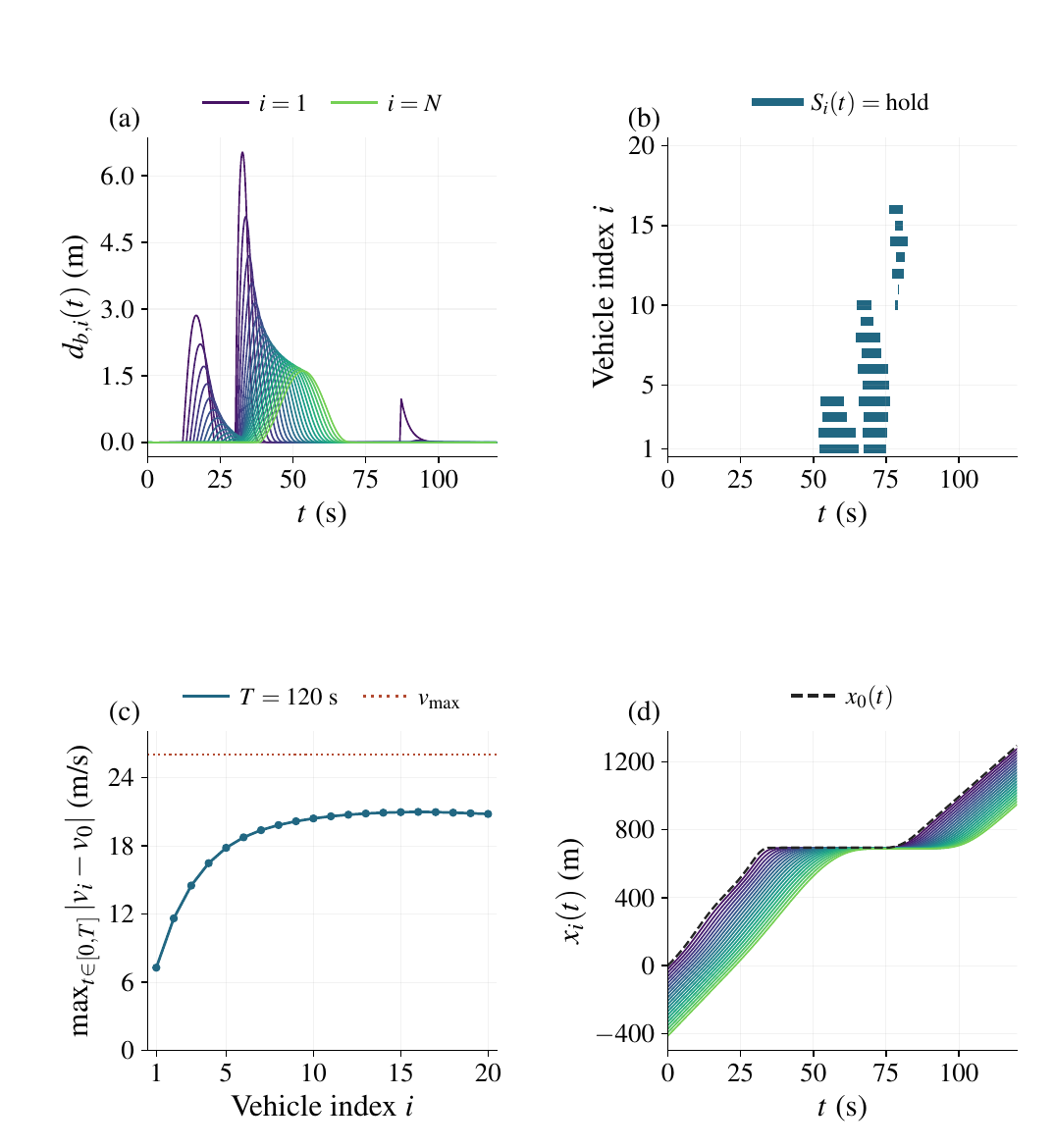}
  \caption{Case~2: braking distances, hold intervals, peak velocity deviations,
  and positions.}
  \label{fig:case2_diagnostics}
\end{figure}

\begin{figure}[t]
  \centering
  \includegraphics[width=\columnwidth]{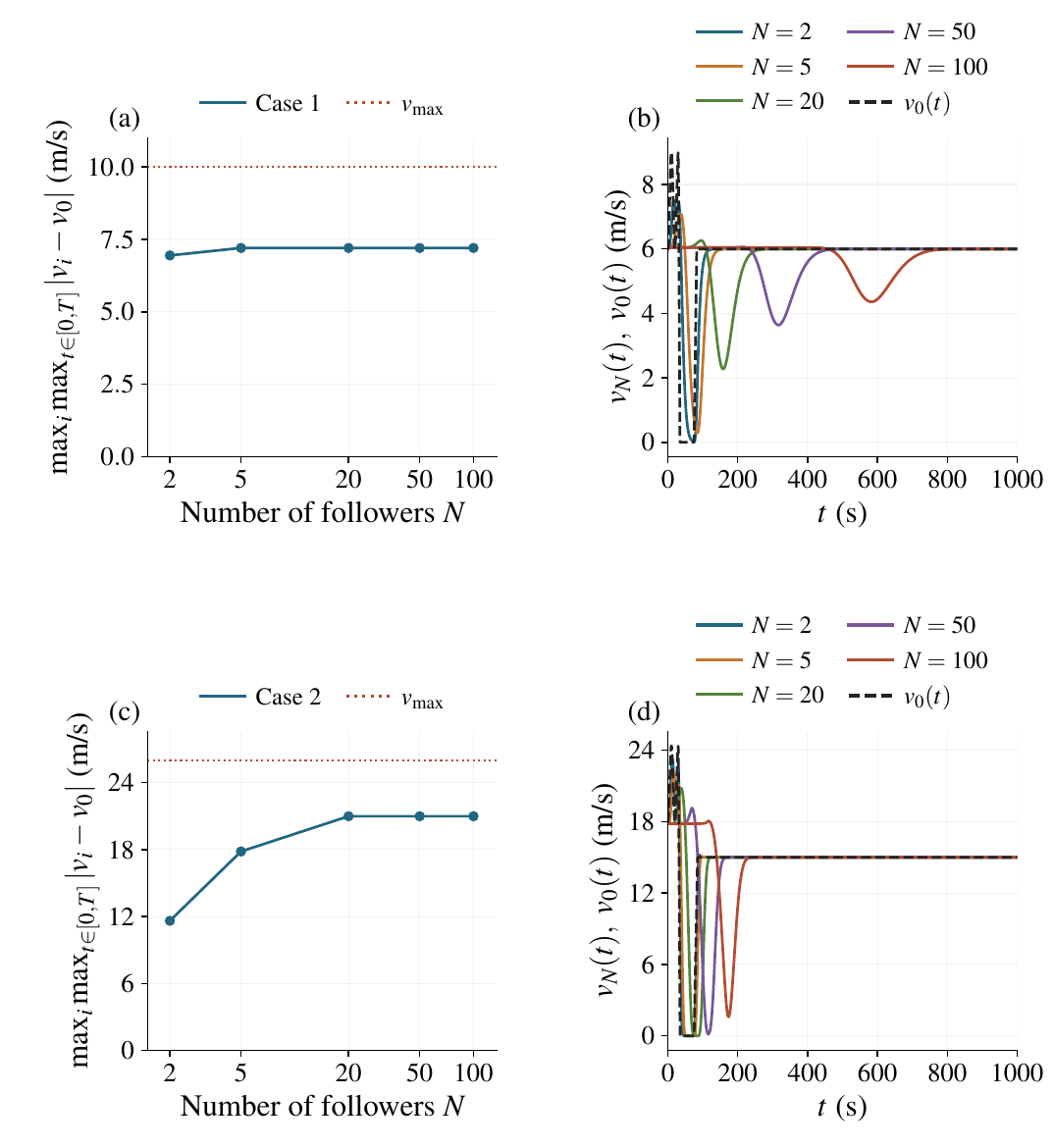}
  \caption{Platoon-length study, $T=1000~\mathrm{s}$.
  Top: Case~1. Bottom: Case~2.}
  \label{fig:length_comparison}
\end{figure}
We use $m_i=1500+(-1)^{i+1}1100~\si{\kilo\gram}$,
$c_d=1$, and propulsion coefficients $0.2$ for heavy vehicles
and $0.8$ for light vehicles.
We set $\theta=0.05~\si{\radian}$, $\bar d=150~\si{\newton}$,
$\vmax=26~\si{\metre\per\second}$, $\mu=0.02$,
$\lambda=1.1~\si{\second}$, and $\gamma=0.1$.
Initially, all vehicles travel at $18~\si{\metre\per\second}$,
with spacing approximately $20.713~\si{\metre}$.
Here $\varpi\approx0.0882$, whereas \ref{P3} requires
$\lambda>385.7~\si{\second}$, so the theorem does not cover this case.

The leader follows
$v_0(t)=18+(10/\pi)(1-\cos(\pi t/10))$ for
$30~\si{\second}$, then brakes at
$5~\si{\metre\per\second\squared}$ to standstill.
At $75~\si{\second}$ it restarts at
$1.5~\si{\metre\per\second\squared}$ to
$15~\si{\metre\per\second}$.
Over $T=120~\si{\second}$, the minimum gap is approximately
$0.304~\si{\metre}$ and the errors remain inside their funnels.
During restart, the heavy vehicles experience sustained
propulsion saturation. The first follower's lower boundary
relaxes to approximately $-4.41~\si{\metre}$ and then returns
toward its nominal value (Fig.~\ref{fig:case2_trajectories}).
The standstill mechanism activates and releases for
$16$ followers (Fig.~\ref{fig:case2_diagnostics}).

\emph{Effect of platoon length.}
We repeat both scenarios for $N\in\{2,5,20,50,100\}$ over
$T=1000~\si{\second}$, keeping each vehicle's parameters and
disturbance phase fixed when followers are added.
The maximum leader-relative velocity deviation is approximately
$7.20~\si{\metre\per\second}$ for $N\ge5$ in Case~1 and
$20.99~\si{\metre\per\second}$ for $N\ge20$ in Case~2
(Fig.~\ref{fig:length_comparison}).
The tail response becomes more delayed as $N$ grows.
These observations concern the tested lengths and horizons;
they do not establish disturbance attenuation along arbitrary platoons.

\emph{Emergency-braking comparison.}
\begin{figure}[t]
  \centering
  \includegraphics[width=\columnwidth]{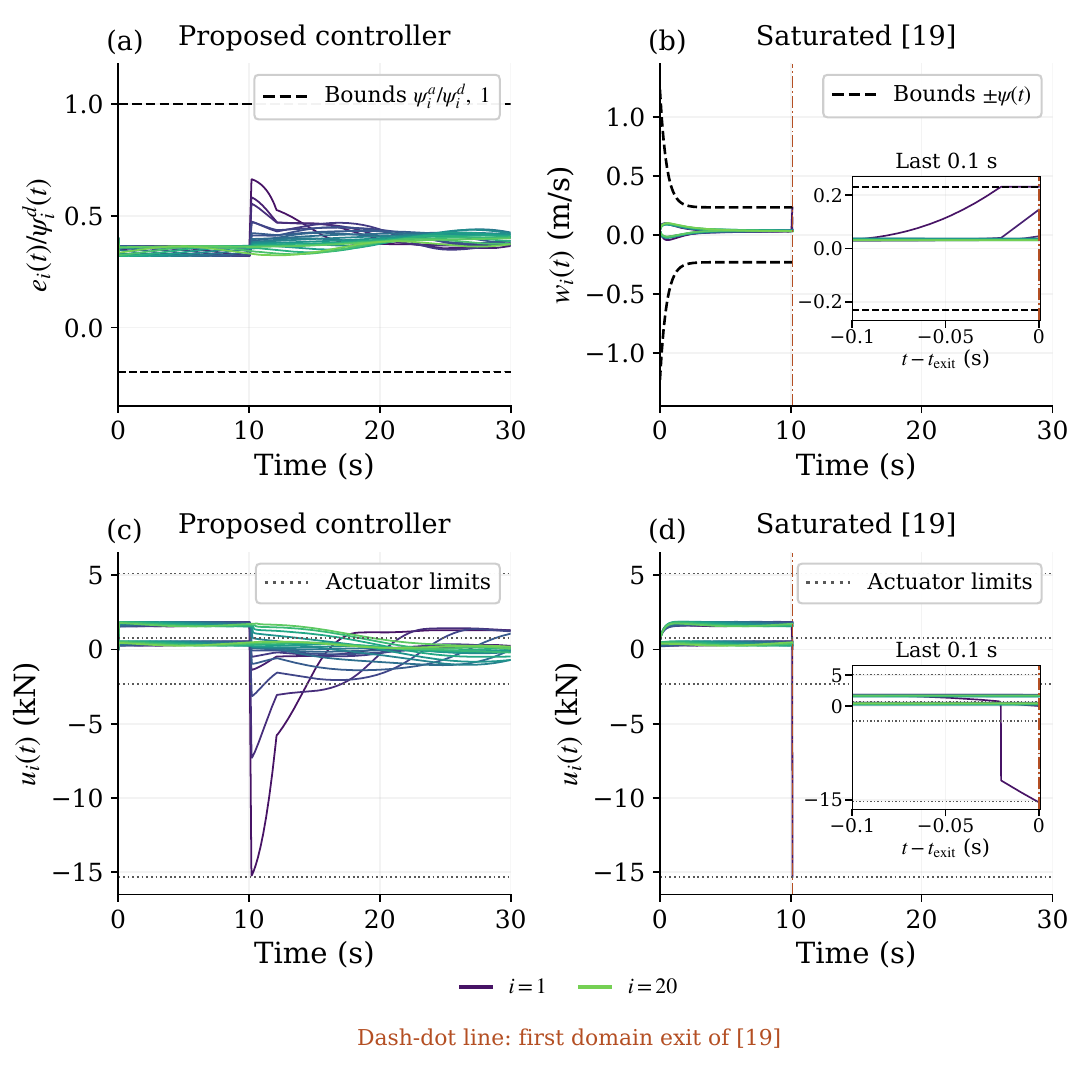}
  \caption{Emergency braking: proposed controller (a),(c) and
saturated controller of~\cite{BergBess26} (b),(d).
Top: errors $e_i/\psi_i^d$ (a) and $w_i$ (b), with funnel
bounds (dashed). Bottom: applied forces and actuator limits
(dotted). Insets in (b),(d): final $0.1~\si{\second}$ before
termination. Colors identify followers.}
  \label{fig:braking_comparison}
\end{figure}

We compare two separate platoons with identical vehicle parameters,
initial conditions, disturbances, leader motion, and actuator limits.
In Fig.~\ref{fig:braking_comparison}, ``Proposed controller'' means
that all followers use our controller~\eqref{eq:SSFCC};
``Saturated~\cite{BergBess26}'' means that all followers use the
controller of~\cite{BergBess26}, with its commanded force saturated
at the same actuator limits.
In either platoon, the first follower is vehicle $i=1$,
immediately behind the leader $i=0$.

We use the Case~2 masses, road, disturbances, and controller settings,
but set $c_{a,i}=0.2$ and $c_d=0.6$ for both implementations.
Initially, all vehicles travel at
$v_{\mathrm{ref}}=18~\si{\metre\per\second}$ with gap
$20.91~\si{\metre}$, giving $\xi_i(0)=0$ for the proposed controller.
At $t=10~\si{\second}$, the leader brakes with peak deceleration
$9~\si{\metre\per\second\squared}$ and $0.2~\si{\second}$ cosine
ramps at the start and end of braking.
It stops at $t=12.2~\si{\second}$ and remains stationary until
$T=200~\si{\second}$.
These parameters do not satisfy~\ref{P3}.

For the controller of~\cite{BergBess26}, we use $\lambda=1.1$,
$k_1=k_2=5200$, $d_{\min}=1~\si{\metre}$,
$d_{\max}=100~\si{\metre}$, and
$\psi(t)=e^{-2t}+\psi_\infty$, where $\psi_\infty=
 \tfrac{\psi_{f,i}^d+\psi_{f,i}^a}{2h'(v_{\mathrm{ref}})}
 \approx0.2323~\si{\metre\per\second}.$ Here, $\psi_{f,i}^a>0$ denotes the magnitude of the nominal lower
boundary. This choice matches the asymptotic half-width to the locally
linearized nominal velocity half-width of the proposed controller
at $v_{\mathrm{ref}}$, holding distance and predecessor velocity fixed.
It does not make the two performance specifications globally equivalent.
Writing $d_i=x_{i-1}-x_i$, the velocity-funnel error
in~\cite{BergBess26} is
\[
 w_i=v_i-v_{i-1}+\frac{1}{d_i-d_{\min}}
                      -\frac{1}{d_{\max}-d_i}.
\]
Thus, this controller includes distance-dependent repulsion even
though it does not use our braking-distance term.
Its original law requires $d_{\min}<d_i<d_{\max}$ and
$|w_i|<\psi(t)$.

Both methods use the same nonnegative-velocity guard and hard force limits. Each simulation stops at $T$ or when any follower reaches its controller's domain boundary, whichever occurs first.
Each simulation stops at $T$ or when any follower reaches its controller's domain boundary, whichever occurs first.
At a funnel boundary, the controller becomes singular, although the applied force remains bounded by saturation.

For $N=20$, the saturated controller of~\cite{BergBess26} reaches its
upper velocity-funnel boundary at $t\approx10.131~\si{\second}$,
at the first follower ($i=1$).
The smallest gap across its platoon up to termination is approximately
$20.770~\si{\metre}$, so no collision is observed before termination.
The proposed controller completes the $T=200~\si{\second}$ simulation
with all errors inside their funnels and a minimum gap of approximately
$0.303~\si{\metre}$.
All adaptive lower boundaries remain at
$\psi_i^a=-0.1~\si{\metre}$, so no funnel relaxation occurs
in this comparison.

Figure~\ref{fig:braking_comparison} displays all $20$ followers
over the first $30~\si{\second}$.
The normalized bounds in panel~(a) remain at
$\psi_i^a/\psi_i^d=-0.2$ and $1$.
The insets in panels~(b),(d) show the final $0.1~\si{\second}$
before termination of the saturated controller's simulation,
with time measured relative to termination time.
The follower colors match Cases~1 and~2, and the shared legend
applies to every panel and inset.
We omit full-horizon force-effort and peak-response comparisons
because the two simulations have different durations.

\emph{Acceleration and funnel adaptation.}
\begin{figure}[t]
  \centering
  \includegraphics[width=\columnwidth]{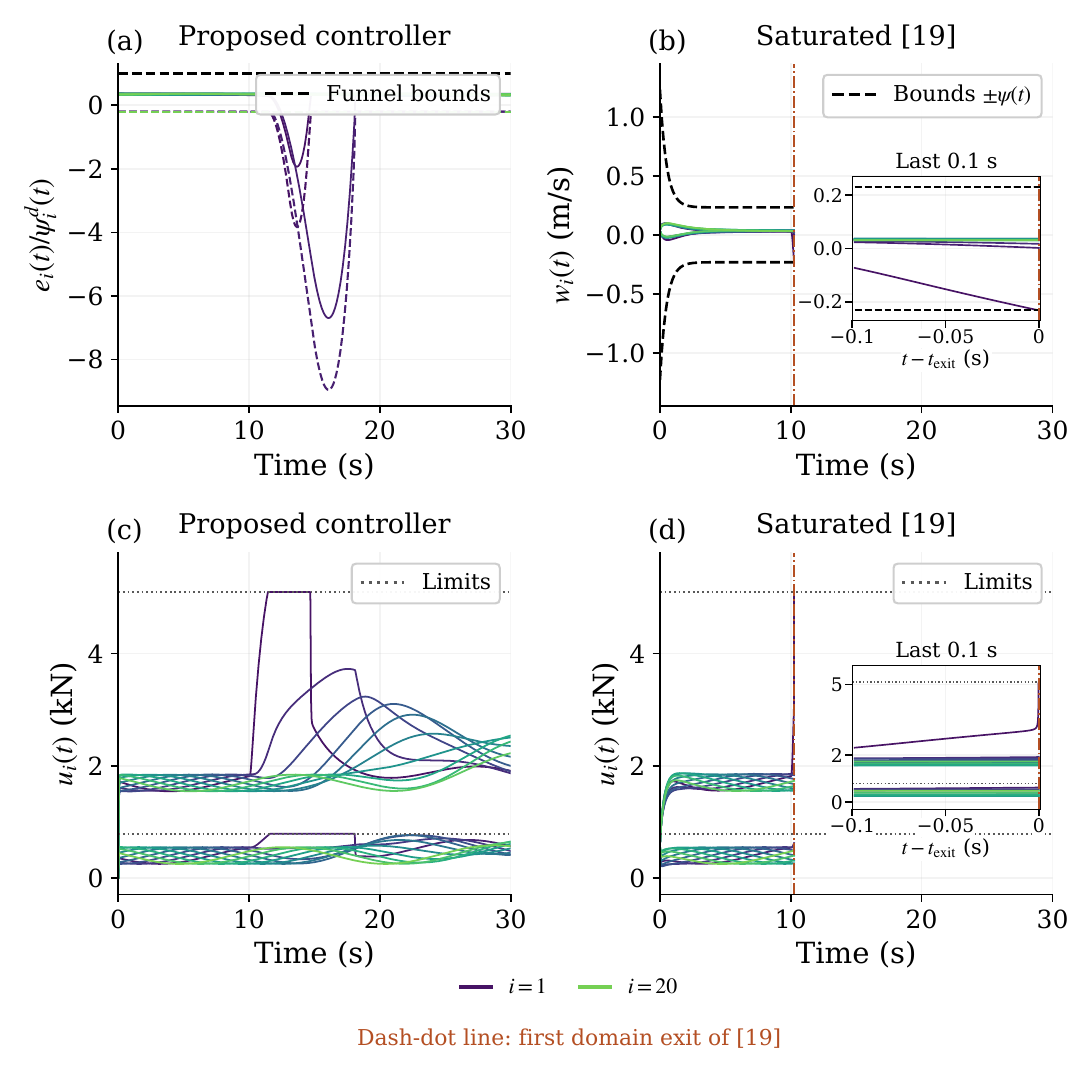}
  \caption{Acceleration: proposed controller (a),(c) and
saturated controller of~\cite{BergBess26} (b),(d).
Top: errors $e_i/\psi_i^d$ (a) and $w_i$ (b), with funnel
bounds (dashed; adaptive in (a)). Bottom: applied forces and
actuator limits (dotted). Insets in (b),(d): final
$0.1~\si{\second}$ before termination. Colors identify followers.}
  \label{fig:acceleration_comparison}
\end{figure}

We retain the same $N=20$ followers, initial conditions, controller
parameters, disturbances, and actuator limits as in the braking
comparison, changing only the leader maneuver.
Starting at $t=10~\si{\second}$, the leader accelerates from
$18$ to $24~\si{\metre\per\second}$ with a prescribed peak acceleration
of $2~\si{\metre\per\second\squared}$.
Cosine ramps of $0.2~\si{\second}$ connect the constant-acceleration
interval to cruising, and the leader reaches its final speed at
$t=13.2~\si{\second}$.
This peak acceleration is a specified scenario limit for the exogenous
leader, rather than a physical limit derived from a leader vehicle model.

The saturated controller of~\cite{BergBess26} reaches its lower
velocity-funnel boundary at $t\approx10.251~\si{\second}$,
at the first follower ($i=1$).
At termination, this follower travels approximately
$0.269~\si{\metre\per\second}$ slower than the leader.
As $w_1+\psi(t)\to0^+$, its unsaturated force command tends to
$+\infty$, while the applied force remains bounded by the actuator limit.
The simulation terminates at this domain boundary and is not continued
through the singularity. All recorded inter-vehicle gaps remain positive.

The proposed controller completes the $T=200~\si{\second}$ simulation
with all errors inside their adaptive funnels.
During propulsion saturation, the lower bounds relax, with the smallest
value $\psi_2^a\approx-4.478~\si{\metre}$ attained at
$t\approx16.08~\si{\second}$.
The bounds subsequently recover toward their nominal value
$-0.1~\si{\metre}$ (Fig.~\ref{fig:acceleration_comparison}).
With adaptation disabled ($\gamma=0$) and all other settings unchanged,
vehicle $i=1$ reaches its fixed lower boundary at
$t\approx12.304~\si{\second}$.
Thus, in this test, adaptation accommodates the temporary tracking error
during propulsion saturation, allowing the constrained controller to
remain well defined.

\section{Conclusion}\label{Sec:Concl}
We proved that safety and link-wise velocity string stability can be ensured under actuation constraints, including standstill situation, i.e., every follower tracks the
velocity range of its predecessor with gain one up to an explicit offset, while
speeds, spacing errors and control signals are bounded uniformly in the platoon
length. The simulations showed that the sufficient headway condition is
conservative, i.e., substantially smaller headways also perform well in
the tested scenarios. Future work will seek less conservative safety conditions and
replace the assumed knowledge of the preceding vehicle's velocity
with estimates obtained from range measurements, e.g., LiDAR,
or visual feedback, e.g., cameras.

\section*{Declaration on the use of AI-assisted tools}
OpenAI ChatGPT and Anthropic Claude were used to assist with language editing of the manuscript and with the development and debugging of parts of the numerical implementation.  All mathematical derivations, numerical results, and scientific conclusions are those of the authors. The authors independently verified all AI-assisted code and reviewed all AI-assisted content, and remain fully responsible for the content of the manuscript.
\appendices
\section{Proof of Theorem~\ref{thm:main}}
\label{app:proof-main}
\begin{proof}
The proof is divided into several steps.

\textit{Step 1: Local existence of a maximal switching solution.}
For each \(i\in\{1,\dots,N\}\), let \(\bar\psi_i^a>0\) to be fixed later, with
\[
-\bar\psi_i^a<\psi_i^{a,0}<0 .
\]
The closed-loop dynamics form a switched system $\dot z(t)=F_{\mathcal S(t)}(t,z(t)) $, with $z(t) = (z_1(t),\dots,z_N(t))^\top \in\mathcal X $, $z_i(t)=(x_i(t),v_i(t),\psi_i^a(t))^\top$,
where, for each $i=1,\dots,N$, the corresponding dynamics is
\[
\dot z_i(t)=
\begin{pmatrix}
v_i(t)\\[4pt]
\begin{cases}
\dfrac{1}{m_i}
\varphi_i\bigl(t,x_i(t),v_i(t),u_i(t)\bigr),
& \mathcal S_i(t)=\mathrm{drive},\\[4pt]
0,
& \mathcal S_i(t)=\mathrm{hold},
\end{cases}\\[10pt]
\begin{aligned}
&-\ell_i^a\bigl(\psi_i^a(t)+\psi_{f,i}^a\bigr)\\
&\quad
+s_\gamma\bigl(-\xi_i(t)\bigr)
\frac{u_i(t)-u_{d_i}(t)}{1-\xi_i(t)}
\end{aligned}
\end{pmatrix}.
\]
To express the closed-loop vector field compactly, define the function
\[
    h:(-\infty,\vmax)\to\R,\ v\mapsto \lambda v + \frac{\mu v^2}{\vmax-v}
\]
and the sets
\[
\mathcal D_i:=
\left\{
\begin{aligned}
&(z,v,\hat v,\psi)\in\R^4:\quad \psi<0,\ v<\vmax\\
&\psi<
z+\sigma_a\!\left(
\frac{v^2-\hat v^2}
     {2g(\underline c_d-\varpi)}
\right)
+h(v)+\psi^d_{f,i}
<\psi^d_{f,i}
\end{aligned}
\right\}.
\]
Here, $z$, $v$, $\hat v$, and $\psi$ represent the relative
position, ego velocity, predecessor velocity, and adaptive lower
funnel boundary, respectively.

The transformed-error map
$\xi_i:\mathcal D_i\to(-1,1)$ is given by
\begin{align*}
\xi_i(z,v,\hat v,\psi)
&:=
1+\frac{2}{\psi^d_{f,i}-\psi}
\bigg(
z+h(v)\\
&\hspace{3.8em}
+\sigma_a\!\left(
\frac{v^2-\hat v^2}
     {2g(\underline c_d-\varpi)}
\right)
\bigg),
\end{align*}
and the feedback map $C_i:\mathcal D_i\to\R$ is
\begin{align*}
C_i(z,v,\hat v,\psi)
&:=
-\frac{k_i}
{1-\xi_i(z,v,\hat v,\psi)^2}
\ln\!\left(
\frac{1+\xi_i(z,v,\hat v,\psi)}
     {1-\xi_i(z,v,\hat v,\psi)}
\right)\!.
\end{align*}
According to~\eqref{eq:SSFCC}, the corresponding closed-loop
signals satisfy
\begin{align*}
 u_{i}(t) &= \mathrm{sat}_{\underline{u}_i}^{\bar{u}_i}\!\big(u_{d_i}(t)\big),\\
 u_{d_i}(t) &=
C_i\bigl(x_i(t)-x_{i-1}(t),v_i(t),v_{i-1}(t),\psi_i^a(t)\bigr),\\
\xi_i(t)
&=
\xi_i\bigl(x_i(t)-x_{i-1}(t),v_i(t),v_{i-1}(t),\psi_i^a(t)\bigr).
\end{align*}

Consequently, for every mode vector
$\mathcal S\in\{\mathrm{drive},\mathrm{hold}\}^N$, the
right-hand side $F_{\mathcal S}$ is defined on
\[
\mathcal X:=
\left\{
\begin{aligned}
&(t,x,v,\psi)\in
\R_{\ge0}\times\R^{2N}
\times\bigtimes_{i=1}^N(-\bar\psi_i^a,0):\\
&\bigl(x_i-x_{i-1},v_i,v_{i-1},\psi_i^a\bigr)
\in\mathcal D_i,\quad i=1,\ldots,N
\end{aligned}
\right\}.
\]
For $i=1$, the quantities $x_{i-1}$ and $v_{i-1}$ in this
definition denote $x_0(t)$ and $v_0(t)$, respectively. The set
$\mathcal X$ is relatively open in
$\R_{\ge0}\times\R^{3N}$.

Because of the standstill mechanism in Assumption~\ref{ass:standstill}, the
closed-loop right-hand side is mode-dependent. The mode vector
$\mathcal S(t)=(\mathcal S_1(t),\dots,\mathcal S_N(t))$ is a piecewise-constant,
time-dependent signal determined by the guards. On each interval on which it is
constant, equal to a fixed value $\mathcal S\in\{\mathrm{drive},\mathrm{hold}\}^N$,
the dynamics are governed by a vector field \(F_{\mathcal S}(t,z)\), where the
velocity component of vehicle \(i\) follows the force-balance model
\eqref{eq:Sys} when \(\mathcal S_i=\mathrm{drive}\), and satisfies
\(\dot v_i=0\) when \(\mathcal S_i=\mathrm{hold}\). Thus the closed-loop system
is interpreted as a switched system, solved mode by mode.

On every interval on which \(\mathcal S\) is constant, the active vector field
\(F_{\mathcal S}\) satisfies the conditions of the Carathéodory local
existence and uniqueness theorem \cite[\S10]{Walter1998} on compact subsets of \(\mathcal X\). Indeed, \(F_{\mathcal S}\) is a Carathéodory
function: it is measurable in \(t\) (the disturbances \(d_i\) being essentially
bounded and measurable) and locally Lipschitz in the closed-loop state: the
functions \(f_i\), \(\sigma_a\), and
\(\mathrm{sat}_{\underline u_i}^{\bar u_i}(\cdot)\) are locally Lipschitz, cf.\ Assumption~\ref{ass:bounds}, and $h$ is smooth on
$(-\infty,\vmax)$ with both $h$ and $h'$
being bounded on any compact subset, thus $h$ is locally Lipschitz as well. Moreover, the denominators $\psi_{f,i}^d-\psi_i^a(t)$, $1-\xi_i(t)$, $1-\xi_i(t)^2$, $\vmax-v_i(t)$
stay strictly away from zero on compact subsets of \(\mathcal X\), and the frozen components
in hold mode are trivially Lipschitz. Hence there exists a unique absolutely
continuous solution on each fixed-mode interval.

The switching solution is obtained by concatenating these solution segments at
the switching times of the standstill mechanism.  The initial mode is consistent with the guards by Assumption~\ref{ass:standstill}. Since the guards are deterministic and the standstill switching further has positive
dwell time by Assumption~\ref{ass:standstill}, the switching times cannot
accumulate (cf.\ Remark~\ref{zeno}); hence, since
$(0,z(0))\in \mathcal X$ there exists a maximal switched solution
\[
z:[0,\omega)\to\R^{3N},
~
\omega\in(0,\infty],~\forall\, t\in [0,\omega):\ (t,z(t))\in \mathcal X
\]
which is locally absolutely continuous, since the right-hand side of the switched differential equation~\eqref{eq:standstill} is continuous in $z$ and measurable in $t$. That $(0,z(0))\in\mathcal X$ holds is guaranteed by \ref{P2}, \ref{P4}
and \ref{P5}: the former two give $v_i(0)\in[0,\vmax)$ and $G_i>0$, and the
latter gives $|\xi_i(0)|<\bar\xi^{\,0}_i<1$ by \eqref{eq:xi0}.

By the continuation property of the local theorem \cite[\S10]{Walter1998}, the maximal switched solution can fail to extend only if the closure of its
graph $\{(t,z(t)) : t\in[0,\omega)\}$ is not a compact subset of
$\mathcal X$; across modes, the only additional obstruction would
be accumulation of switching times, which is excluded above. Thus, to prove
forward completeness, i.e., $\omega=\infty$, it suffices to show that this graph
closure would be compact whenever $\omega<\infty$.

\textit{Step 2: Nonnegative velocity.} Fix \(i\in\{1,\dots,N\}\) and $t\in [0,\omega)$. If \(\mathcal S_i(t)=\mathrm{hold}\), then
\(v_i(t)=0\). Consider now the boundary \(v_i(t)=0\) in drive mode. If
\(\varphi_i(t)\le0\), the guard \(\mathrm{drive}\to\mathrm{hold}\) is activated and
the hold mode enforces \(\dot v_i(t)=0\). If $\varphi_i(t)>0$, the guard
$\mathrm{hold}\to\mathrm{drive}$ keeps the vehicle in drive mode and
$\dot v_i(t)=\frac{\varphi_i(t)}{m_i}>0$.
Thus, at \(v_i(t)=0\), the effective vector field is either tangent to the boundary
of \(\mathbb R_{\ge0}\) or points into \(\mathbb R_{>0}\). Therefore no solution
can cross from \(\mathbb R_{\ge0}\) into \(\mathbb R_{<0}\). Since
\(v_i(0)\ge0\), it follows that \(v_i(t)\ge0\) for all \(t\in [0,\omega)\) in the interval of
existence.

\emph{Step 3: Uniform bounds on velocity and acceleration.}
By Step~2 and by the definition of $\mathcal X$ we have
$v_i(t)\in[0,\vmax)$ for all $t\in[0,\omega)$ and $i=1,\dots,N$. The estimates below are
first derived in drive mode, where the force-balance model applies; in hold
mode one has $v_i=0$ and $\dot v_i=0$, so the bounds are valid in both modes.\\
Since $u_i(t)\le\bar u_i=c_{a,i}m_ig$, $v_i(t)\ge0$, Assumption~\ref{ass:bounds}
holds, and the drag contribution $-\frac{\underline c_\rho}{m_i}v_i(t)^{2}$ is
non-positive, we obtain
\[
  \dot v_i(t)\le\dot v^{u}
  :=g(\bar c_a+\sin\bar\theta)+\frac{\bar d}{\underline m},
\]
where the rolling friction is omitted: for $v_i(t)\ge0$ it is retarding and,
under the smooth approximation, $s_\kappa(v_i)\to0$ as $v_i\to0$, so it cannot
be credited in an upper bound. On the other hand, since
$u_i(t)\ge-c_{d,i}m_ig\ge-\bar c_dm_ig$ and $0\le v_i(t)<\vmax$,
\[
  \dot v_i(t)\ge\dot v^{\ell}
  :=-g(\bar c_d+\bar C_r+\sin\bar\theta)-\frac{\bar d}{\underline m}
    -\frac{\bar c_\rho}{\underline m}\vmax^{2}.
\]
By Remark~\ref{rem:collapse}, $\bar a$ from \ref{P2} satisfies
$\bar a=\max\{|\dot v^{\ell}|,|\dot v^{u}|\}$, hence $|\dot v_i(t)|\le\bar a$
for all $t\in[0,\omega)$ and all $i=1,\dots,N$.

\textit{Step 4: Uniform bounds on $\epsilon_i$, $\xi_i$, $\zeta_i$ and $\psi_i^a$.} From the closed-loop dynamics and~\eqref{eq:SSFCC} we obtain, for all $t\in [0,\omega)$:
\begin{align}
\dot\xi_i(t)
={}&
\frac{2}{\psi^d_{f,i}-\psi_i^a(t)}
\Bigg(
\dot e_i(t) \notag+\frac{1-\xi_i(t)}{2}\ell_i^a
\bigl(\psi_i^a(t)+\psi^a_{f,i}\bigr) \notag\\
&+\frac{1}{2}s_\gamma\bigl(-\xi_i(t)\bigr)
\bigl(u_{d_i}(t)-u_i(t)\bigr)
\Bigg),\notag\\
\dot e_i(t)
={}&v_i(t)-v_{i-1}(t) +\sigma_a'\bigl(\varsigma_i(t)\bigr)
\dot\varsigma_i(t)
+h'\bigl(v_i(t)\bigr)\dot v_i(t),
\label{epsdot}
\end{align}
where $\varsigma_i(t) = \frac{v_i(t)^2 - v_{i-1}(t)^2}{2g(c_d - \varpi)}$, with derivative
\begin{equation}\label{eq:varsigmadot}
\dot\varsigma_i(t)
=
\frac{v_i(t)\dot v_i(t)-v_{i-1}(t)\dot v_{i-1}(t)}{g(\underline c_d-\varpi)} .
\end{equation}
Consider $V_i(t):=\tfrac12\epsilon_i(t)^2$ for $t\in[0,\omega)$.
Then
\begin{equation}\label{eq:Vdot_step3_compact}
\begin{aligned}
\dot V_i(t)
={}&
\frac{4\zeta_i(t)\epsilon_i(t)}
{\psi^d_{f,i}-\psi_i^a(t)}
\Bigg(
\dot e_i(t)+\frac{1-\xi_i(t)}{2}\ell_i^a
\bigl(\psi_i^a(t)+\psi^a_{f,i}\bigr)\\
&+\frac{1}{2}s_\gamma\bigl(-\xi_i(t)\bigr)
\bigl(u_{d_i}(t)-u_i(t)\bigr)
\Bigg).
\end{aligned}
\end{equation}
Before proceeding, we show that
\(\psi_i^a(t)+\psi^a_{f,i}\le0\) for all \(t\in[0,\omega)\). By assumption,
\(\psi_i^a(0)=-\psi^a_{f,i}\). Consider the boundary
\(\psi_i^a=-\psi^a_{f,i}\) (at an arbitrary time). At this boundary, the first term in the dynamics of
\(\psi_i^a\) vanishes. Moreover, if \(\xi_i>0\), then \(s_\gamma(-\xi_i)=0\)
(and the boundary case $\xi_i=0$ gives $\epsilon_i=0$, hence $u_{d_i}=u_i$, so
the term vanishes as well).
If \(\xi_i<0\), then \(\epsilon_i<0\), and hence
\(u_{d_i}=-k_i\zeta_i\epsilon_i>0\). Since \(u_{i}\) is the saturation of
\(u_{d_i}\), we have \(u_{i}\le u_{d_i}\). Therefore, $s_\gamma(-\xi_i)\frac{u_{i}-u_{d_i}}{1-\xi_i}\le0 $. Consequently, \(\dot\psi_i^a\le0\) whenever
\(\psi_i^a=-\psi^a_{f,i}\). It follows that $\psi_i^a(t)\le-\psi^a_{f,i}$ for all $t\in[0,\omega)$.

\emph{Case 1: $\xi_i(t)\ge 0$.}
Then $\epsilon_i(t)\ge 0$, $s_\gamma(-\xi_i(t))=0$, and since
$\psi^a_i(t)+\psi^a_{f,i}\le 0$:
\[
\dot V_i(t)
\le
\frac{4\zeta_i(t)\epsilon_i(t)}{\psi^d_{f,i}-\psi_i^a(t)}\,\dot e_i(t),
\qquad
\frac{4\zeta_i(t)\epsilon_i(t)}{\psi^d_{f,i}-\psi_i^a(t)}\ge 0 .
\]
It therefore suffices to control the sign of $\dot e_i$ as $\xi_i$ approaches
the upper boundary at $\xi_i=1$. We treat the drive and hold modes separately.

\noindent\emph{Drive mode.} For brevity we omit arguments where they are clear from
the context, and abbreviate $\delta:=g(\underline c_d-\varpi)$. In drive mode we have $\dot v_i=\tfrac1{m_i}(u_i-f_i+d_i)$. Inserting
\eqref{eq:varsigmadot} into \eqref{epsdot} and collecting the terms
containing $\dot v_i$ gives
\[
  \dot e_i=(v_i-v_{i-1})
    -\sigma_a'(\varsigma_i)\frac{v_{i-1}\dot v_{i-1}}{\delta}
    +\Bigl(\sigma_a'(\varsigma_i)\frac{v_i}{\delta}+h'(v_i)\Bigr)\dot v_i .
\]
Since $\sigma_a'\ge0$, $v_i\ge0$ and $h'(v_i)\ge\lambda>0$, the coefficient of
$\dot v_i$ is non-negative. As $\xi_i\to1$ one has $\epsilon_i\to+\infty$ and
thus $u_{d_i}=-k_i\zeta_i\epsilon_i\to-\infty$, so there is
$\bar\xi_i<1$ with $u_i=\mathrm{sat}^{\bar u_i}_{\underline u_i}(u_{d_i})
=-c_{d,i}m_ig$ for all $\xi_i\ge\bar\xi_i$; for such $\xi_i$,
$\dot v_i\le\frac{u_i}{m_i}+g\varpi=-g(c_{d,i}-\varpi)\le-\delta$. Using
$\sigma_a'\le1$, $v_{i-1}\le\max\{\vmax,\bar v_0\}=\vmax$,
$|\dot v_{i-1}|\le\max\{\bar a,\bar a_0\}$ and $v_i\le \vmax$ (by Step~3) we infer
\[
  \dot e_i\le
   \vmax +\frac{\vmax\max\{\bar a,\bar a_0\}}{\delta}-\delta\,\lambda,
\]
Since $\lambda>\beta /\underline\alpha$ by \ref{P3} and 
\[
  \underline\alpha \le g\underline c_a \le g\bar c_a
  \le g(\underline c_d-\varpi) =\delta
\]
by \ref{P1}, it follows that $\lambda >\beta/\delta = \frac{\vmax}{\delta}+\frac{\vmax\max\{\bar a,\bar a_0\}}{\delta^{2}}$, thus $\dot e_i\le0$
and hence $\dot V_i\le0$ for all $\xi_i\ge\bar\xi_i$, so $\xi_i$ cannot cross
the upper boundary in drive mode.

\noindent\emph{Hold mode.}
It remains to verify that the hold mode does not disrupt the
drive-mode argument. In hold mode we have $v_i = 0$ and $\dot v_i=0$, so
\[
\varsigma_i
=
\frac{v_i^2-v_{i-1}^2}{2g(\underline c_d-\varpi)}
=
-\frac{v_{i-1}^2}{2g(\underline c_d-\varpi)}
\le0,
\]
and therefore, by the one-sided property of the regularization,
$d_{b,i}=\sigma_a(\varsigma_i)=0$ with $\dot d_{b,i}=0$. Since moreover $h(0)=0$ and $\dot v_i=0$, we obtain
\[
  e_i=x_i-x_{i-1}+\psi^{d}_{f,i},\qquad
  \dot e_i=v_i-v_{i-1}=-v_{i-1}\le0,
\]

the last inequality following from Step~2.
Since $\xi_i\ge0$ (by assumption of Case~1), we have  $\tfrac{4\zeta_i\epsilon_i}{\psi^d_{f,i}-\psi_i^a}\ge0$, and using
$s_\gamma(-\xi_i)=0$ together with $\psi_i^a+\psi^a_{f,i}\le0$, it follows that
\[
\dot V_i
=
\frac{4\zeta_i\epsilon_i}{\psi^d_{f,i}-\psi_i^a}
\left(
\dot e_i
+
\frac{1-\xi_i}{2}\ell_i^a(\psi_i^a+\psi^a_{f,i})
\right)
\le0 .
\]
Hence, exactly as in drive mode, $V_i$ does not increase in hold mode, so $\xi_i$
stays uniformly away from the upper boundary $\xi_i=1$. Consequently, hold mode
cannot cause an escape through the upper boundary, and a possible increase
of $e_i$ toward that boundary needs only be excluded during drive mode, and is guaranteed under \ref{P1} and \ref{P3}.

\emph{Case 2: \(\xi_i(t)<0\).}
Then \(\epsilon_i(t)<0\) and $|\zeta_i(t)\epsilon_i(t)|=-\zeta_i(t)\epsilon_i(t)$. Again, we omit arguments in the following for brevity. As \(\xi_i\to-1\), one has \(\epsilon_i\to-\infty\), and hence $u_{d_i}=-k_i\zeta_i\epsilon_i\to+\infty .$
Thus the acceleration input saturates, $u_i=\bar u_i=c_{a,i} m_i g$, for
$\xi_i\le\underline\xi_i$, where $\underline\xi_i\in(-1,0)$ is made explicit below. Next, invoking~\eqref{eq:Vdot_step3_compact} we get:
\[
\begin{aligned}
\dot V_i
={}&-\frac{4|\zeta_i\epsilon_i|}
{\psi^d_{f,i}-\psi_i^a}
\Bigg(
\dot e_i
+\frac{1-\xi_i}{2}\ell_i^a
\bigl(\psi_i^a+\psi^a_{f,i}\bigr)\\
&\hspace{5.5em}
+\frac{1}{2}s_\gamma(-\xi_i)
\bigl(u_{d_i}-u_i\bigr)
\Bigg).
\end{aligned}
\]
For $\xi_i\le\underline\xi_i$, maximum propulsion is active and $u_i=\bar u_i=c_{a,i}m_i g.$ Therefore, by \ref{P2} and since
$v_i(t)\in[0,\vmax)$ by Step~3,
\begin{align*}
  \dot v_i&=\frac{1}{m_i}\bigl(\bar u_i-f_i(t,x_i,v_i)+d_i\bigr)\\
  &\ge\ g\bigl(\underline c_a-\sin\bar\theta-\bar C_r\bigr)
         -\frac{\bar d+\bar c_\rho\vmax^{2}}{\underline m}
   = \underline\alpha>0,
\end{align*}
uniformly in $i$ and $N$.
Hence, 
under $u_i=\bar u_i$ the
net force satisfies $\varphi_i=m_i\dot v_i\ge m_i\underline\alpha\ge\underline m\,\underline\alpha>0$,
so the hold$\to$drive guard $\varphi_i>0$ is met and the vehicle switches from hold to drive.  Moreover,
using the same estimates on the velocity and braking-distance terms as in
Case~1, and using that, since $v_i\ge0$, the velocity-difference term
satisfies $-v_i+v_{i-1}\le v_{i-1}\le\vmax$, we have
\[
-v_i+v_{i-1}-\dot d_{b,i}\le \vmax\Bigl(1+\frac{\max\{\bar a,\bar a_0\}}{g(\underline c_d-\varpi)}\Bigr)=\beta.
\]
Since
\[
-\dot e_i
=
-v_i+v_{i-1}-\dot d_{b,i}-h'(v_i)\dot v_i
\]
and
$h'(v_i)\dot v_i\ge\lambda\underline\alpha$, it follows that
\[
-\dot e_i\le \beta-\lambda\underline\alpha .
\]
Moreover, $u_{d_i}=-k_i\zeta_i\epsilon_i=k_i|\zeta_i\epsilon_i|$
and therefore
\[
-\frac12|\zeta_i\epsilon_i|\,s_\gamma(-\xi_i)(u_{d_i}-u_i)
\le
-\frac{\gamma k_i}{2}|\zeta_i\epsilon_i|^2
+
\frac{\gamma c_{a,i} m_i g}{2}|\zeta_i\epsilon_i|.
\]
Since the graph of the solution is contained in $\mathcal X$, we have $\psi_i^a > -\bar\psi_i^a$,
hence $-(\psi_i^a+\psi^a_{f,i})<\bar\psi_i^a-\psi^a_{f,i}$; together with
$\tfrac{1-\xi_i}{2}\le1$, we obtain
\[
\begin{aligned}
\dot V_i
\le{}&
\frac{4|\zeta_i\epsilon_i|}
{\psi^d_{f,i}-\psi_i^a(t)}
\Bigg(
\beta-\lambda\underline\alpha
+\ell_i^a\bigl(\bar\psi_i^a-\psi_{f,i}^a\bigr)\\
&\hspace{4.5em}
+\frac{\gamma c_{a,i}m_i g}{2}
-\frac{\gamma k_i}{2}|\zeta_i\epsilon_i|
\Bigg).
\end{aligned}
\]
Furthermore, consider the function
\[
\mathcal T: (0,\infty)\to(0,\infty),\ r\mapsto r\cosh^2\!\left(\frac r2\right),
\]
which is continuous and strictly
increasing. Note that from $\epsilon_i=\ln\frac{1+\xi_i}{1-\xi_i}=2\,\mathrm{artanh}\,\xi_i$
we have $\xi_i=\tanh(\epsilon_i/2)$, so $1-\xi_i^2=\mathrm{sech}^2(\epsilon_i/2)$ and hence
$\zeta_i=\frac{1}{1-\xi_i^2}=\cosh^2(\epsilon_i/2)>0$. Since $\zeta_i>0$, it follows that
\[
|\zeta_i\epsilon_i|
= \zeta_i|\epsilon_i|=
|\epsilon_i|\cosh^2\!\left(\frac{\epsilon_i}{2}\right)
=
\mathcal T(|\epsilon_i|),
\]
where we used that $\cosh$ is even, so $\cosh^2(\epsilon_i/2)=\cosh^2(|\epsilon_i|/2)$
and hence $|\epsilon_i|\cosh^2(\epsilon_i/2)=|\epsilon_i|\cosh^2(|\epsilon_i|/2)=\mathcal T(|\epsilon_i|)$.
We then see that $\dot V_i\le0$ whenever
\begin{equation}\label{eq:case2_lower_bound}
\gamma k_i\mathcal T(|\epsilon_i|)
\ge
2\beta-2\lambda\underline\alpha+2\ell_i^a(\bar\psi_i^a-\psi_{f,i}^a)+\gamma c_{a,i} m_i g .
\end{equation}
As $\xi_i\to-1$, one has $|\epsilon_i|\to\infty$, hence $\mathcal T(|\epsilon_i|)\to\infty$,
so condition~\eqref{eq:case2_lower_bound} is eventually satisfied;
then $\dot V_i\le0$ keeps $|\epsilon_i|$ bounded, so $\xi_i$ stays uniformly away
from $-1$. The precise threshold $\bar\epsilon_i$ is constructed below.

It remains to ensure that \(\psi_i^a\) does not reach
\(-\bar\psi_i^a\). Let \(r_i^a>0\) be the unique solution of
\[
\mathcal T(r_i^a)=\frac{c_{a,i} m_i g}{k_i}.
\]
The saturation threshold $\underline\xi_i$ of $\xi_i$ introduced above is then
explicit, $\underline\xi_i=-\tanh(r_i^a/2)$, since for $\xi_i<0$ we have
\[
 u_{d_i}=k_i\mathcal T(|\epsilon_i|)\ge c_{a,i} m_i g\iff|\epsilon_i|\ge r_i^a\iff
\xi_i\le\underline\xi_i.
\]
Define
\begin{equation}\label{phia}
    \phi_i^a:=1-\underline\xi_i=1+\tanh\!\left(\frac{r_i^a}{2}\right)\in(1,2). 
\end{equation} 
Then, under acceleration saturation $\xi_i\le\underline\xi_i$, monotonicity directly gives
\[
1-\xi_i\ge 1-\underline\xi_i=\phi_i^a .
\]
As $\psi_i^a\to-\bar\psi_i^a$ (this value is never attained, since $\psi_i^a(t)>-\bar\psi_i^a$ for all $t\in[0,\omega)$ by definition of $\mathcal X$), using the dynamics of
\(\psi_i^a\), \(u_i=c_{a,i} m_i g\), and \(1-\xi_i\ge\phi_i^a\), we get in this limit
\[
\dot\psi_i^a
\ge
\ell_i^a(\bar\psi_i^a-\psi_{f,i}^a)
-
\frac{\gamma}{\phi_i^a}
\left(
k_i\mathcal T(|\epsilon_i|)-c_{a,i} m_i g
\right).
\]
Therefore, the inward-pointing condition
\[
\dot\psi_i^a>0
\qquad
\text{as }\psi_i^a\to-\bar\psi_i^a
\]
(which, prevents $\psi_i^a$ from ever
reaching $-\bar\psi_i^a$) is guaranteed, provided that
\begin{equation}\label{eq:case2_upper_bound}
k_i\mathcal T(|\epsilon_i|)
<
c_{a,i} m_i g+\frac{\phi_i^a}{\gamma}\ell_i^a(\bar\psi_i^a-\psi_{f,i}^a) .
\end{equation}

We finally show that, under \ref{P5}, there exists a constant
\(\bar\epsilon_i>|\epsilon_i(0)|\) such that the lower bound in~\eqref{eq:case2_lower_bound} and the upper bound in~\eqref{eq:case2_upper_bound} are simultaneously satisfied whenever
\(\epsilon_i\le-\bar\epsilon_i\). Equivalently, $\mathcal T(\bar\epsilon_i)$ must
lie in the nonempty interval
\begin{equation}\label{eq:case2_interval}
L_i
\le
\mathcal T(\bar\epsilon_i) <
\frac{c_{a,i} m_i g}{k_i}
+
\frac{\phi_i^a}{\gamma k_i}
\ell_i^a(\bar\psi_i^a-\psi_{f,i}^a),
\end{equation}
where
\begin{equation}\label{eq:case2_Li}
\begin{aligned}
L_i:=
\max\Bigg\{&
\mathcal T\bigl(|\epsilon_i(0)|\bigr),\\
&
\frac{2}{\gamma k_i}
\Big(
\beta-\lambda\underline\alpha
+\ell_i^a\bigl(\bar\psi_i^a-\psi_{f,i}^a\bigr)
\Big)
+\frac{c_{a,i}m_i g}{k_i}
\Bigg\}.
\end{aligned}
\end{equation}
Setting
$\Delta:=\lambda\underline\alpha-\beta>0$ and using
$2-\phi^{a}_i\in(0,1)$, the interval \eqref{eq:case2_interval} is non-empty, if
\begin{align*}
  \bar\psi^{a}_i&<\psi^{a}_{f,i}+\frac{2\Delta}{(2-\phi^{a}_i)\ell^{a}_i},\\
  \bar\psi^{a}_i&>\psi^{a}_{f,i}
    +\frac{\gamma\bigl(k_i\mathcal T(|\epsilon_i(0)|)-c_{a,i}m_ig\bigr)}
          {\phi^{a}_i\ell^{a}_i}.
\end{align*}
In addition, the state space requires $\bar\psi^{a}_i>\psi^{a}_{f,i}$, which is implied by the second inequality whenever its
right-hand side exceeds $\psi^{a}_{f,i}$, and is otherwise the binding lower
bound. Hence a feasible $\bar\psi^{a}_i$ exists exactly when
\begin{equation}\label{eq:feasible}
  \max\Bigl\{0,\
  \frac{\gamma}{\phi^{a}_i}
   \bigl(k_i\mathcal T(|\epsilon_i(0)|)-c_{a,i}m_ig\bigr)\Bigr\}
  \;<\;\frac{2\Delta}{2-\phi^{a}_i},
\end{equation}
where $\ell^{a}_i>0$ has been cancelled. Since $\Delta>0$, \eqref{eq:feasible}
reduces to
$\mathcal T(|\epsilon_i(0)|)<\frac{c_{a,i}m_ig}{k_i}
+\frac{2\Delta\phi^{a}_i}{(2-\phi^{a}_i)\gamma k_i}
=\mathcal T(\bar\epsilon^{\,0}_i)$,
that is, to $|\epsilon_i(0)|<\bar\epsilon^{\,0}_i$, equivalently
$|\xi_i(0)|<\bar\xi^{\,0}_i$, which holds by \ref{P5} and \eqref{eq:xi0}.

Since \(\mathcal T\) is continuous, strictly increasing, and maps
\((0,\infty)\) onto \((0,\infty)\), there exists \(\bar\epsilon_i>|\epsilon_i(0)|\)
satisfying \eqref{eq:case2_interval}. Utilizing \eqref{eq:case2_lower_bound} we get
\(\epsilon_i(t)\ge-\bar\epsilon_i\), $t\in[0,\omega)$, and
\eqref{eq:case2_upper_bound} gives the inward-pointing condition
\(\dot\psi_i^a>0\) as \(\psi_i^a\to-\bar\psi_i^a\); hence
\(\psi_i^a(t)>-\bar\psi_i^a\), $t\in[0,\omega)$.

By \ref{P6}, $c_{a,i}m_i/k_i$ is uniformly bounded, hence so is $r^{a}_i$ and
$2-\phi^{a}_i$ is uniformly bounded away from zero; since in addition
$\ell^{a}_i$, $\psi^{a}_{f,i}$ and $\mathcal T(|\epsilon_i(0)|)$ are uniformly
bounded, the admissible values of $\bar\psi^{a}_i$, and therefore
$\bar\epsilon_i$, can be chosen uniformly bounded. Similarly, the
braking-saturation thresholds in Case~1 are uniformly bounded because
$c_{d,i}m_i/k_i\le\bar c_d\bar m/\inf_{N,i}k_i$.

Finally, the estimates above were derived in drive mode. In hold mode,
\(\dot v_i=0\), and by the one-sided property of \(\sigma_a\), we have \(d_{b,i}=0\). Hence, using $h(0)=0$,
\[
\dot e_i=-v_{i-1},
\qquad
-\dot e_i=v_{i-1}\le \vmax \le\beta,
\]
where $\bar v_0<\vmax$ by \ref{P2}, so $e_i$ may drift toward the lower funnel boundary and $\xi_i\to-1$. However,
as $\xi_i\to-1$ the desired input saturates at maximum propulsion,
$u_{d_i}=-k_i\zeta_i\epsilon_i\to+\infty$, so $u_i=\bar u_i=c_{a,i} m_i g$; the
would-be net force then satisfies, by \ref{P2}
evaluated at $v_i=0$,
\[
\varphi_i=\bar u_i-f_i+d_i\ge m_i\underline\alpha\ge\underline m\,\underline\alpha>0,
\]
which triggers the hold$\to$drive guard. Hence the vehicle
releases to drive mode before $\xi_i$ reaches $-1$, where the drive-mode estimate
above applies. The boundary of the state space $\mathcal X$ can therefore not be reached in hold mode.

\textit{Step 5: Global existence and verification of assertions (i)~--~(iii).}
Combining Steps~3 and 4 yields explicit margins: \(v_i(t)\in[0,\vmax)\), and
there exist constants \(\xi_i^{\max}\in(0,1)\) and \(\eta_i>0\) such that
\[
\begin{aligned}
\xi_i(t)
&\in[-\xi_i^{\max},\xi_i^{\max}]
\subset(-1,1),\\
\psi_i^a(t)
&\in[-\bar\psi_i^a+\eta_i,-\psi_{f,i}^a]
\subset(-\bar\psi_i^a,0).
\end{aligned}
\]
for all \(t\in[0,\omega)\). Denote by
\[
  q_i(t):=x_{i-1}(t)-x_i(t),\qquad i=1,\dots,N,
\]
the inter-vehicle distance, so that $q_i(0)>0$ by \eqref{eq:IC}. Suppose, for
contradiction, that $\omega<\infty$. Since $v_{i-1}(t)\le\vmax$ by Step~3 and
\ref{P2}, and $v_i(t)\ge0$ by Step~2, we have $\dot q_i(t)=v_{i-1}(t)-v_i(t)\le\vmax$
and therefore
\[
  q_i(t) \le \bar q_i:=q_i(0)+\vmax\,\omega < \infty,
  \quad t\in[0,\omega).
\]
On the other hand, $\xi_i(t)<1$ is equivalent to $e_i(t)<\psi^{d}_{f,i}$, thus
\begin{equation}\label{eq:key}
  h\bigl(v_i(t)\bigr)+d_{b,i}(t) < q_i(t),
\end{equation}
and $d_{b,i}(t)\ge0$ gives $h(v_i(t))<\bar q_i$. Since $h$ maps $[0,\vmax)$ strictly increasingly onto $[0,\infty)$, its inverse is defined on
$[0,\infty)$ and we obtain
\[
  v_i(t) \le h^{-1}(\bar q_i) < \vmax,\quad t\in[0,\omega),
\]
i.e., $v_i$ stays uniformly away from $\vmax$. In particular $x_i$ remains
bounded on $[0,\omega)$, and the graph of the solution
$z = (x_i,v_i,\psi_i^a)_{i=1}^N$ evolves in a compact subset
of $\mathcal X$.  As $z$ is a maximal switching solution obtained by
concatenating finitely many absolutely continuous Carathéodory arcs (Zeno excluded),
this contradicts the continuation property of Step~1. Therefore, $\omega=\infty$.

Furthermore, \eqref{eq:key} together with $d_{b,i}(t)\ge0$ and $h(v_i(t))\ge0$ (by Step~2) gives $q_i(t)>0$, that is
$x_i(t)-x_{i-1}(t)<0$ for all $t\ge0$ and all $i=1,\dots,N$, which ensures
collision avoidance, i.e., assertion~(ii).

Finally, $v_i(t)\in[0,\vmax)$ was established in Step~3, and $\vmax$, $\underline\alpha$, $\bar a$, $\beta$ are design constants independent of $i$ and $N$. Together with the uniform bounds of \ref{P6}, the bounds derived
above for $\xi_i$ and $\psi^{a}_i$ can be chosen independently of $N$; in particular $\xi_i$ remains uniformly separated from $-1$ and $1$, hence
$\epsilon_i$ and $\zeta_i$ are uniformly bounded, and since the $k_i$ are
uniformly bounded, $u_{d_i}=-k_i\zeta_i\epsilon_i$ is uniformly bounded as
well. This proves assertion~(iii).

\textit{Step 6: Link-wise velocity string stability: verification of
assertion~(iv).}
Let $W_i:=\psi^d_{f,i}+\sup_{t\ge0}(-\psi_i^a(t))$.
By assertion~(iii), established in Step~5, and \ref{P6},
there exists a constant $W>0$, independent of $i$ and $N$,
such that $W_i\le W$.

Recall from Step~5 that $q_i=x_{i-1}-x_i$
satisfies~\eqref{eq:key}. Moreover, $\psi_i^a(t)<e_i(t)$
and the definition of $e_i$ in~\eqref{eq:SSFCC} give
$q_i(t)<h(v_i(t))+d_{b,i}(t)+\psi^d_{f,i}-\psi_i^a(t)$,
so that altogether
\begin{equation}\label{eq:tube}
\begin{aligned}
\forall\,t\ge0:\quad
h(v_i(t))+d_{b,i}(t)
&<q_i(t)\\
&<h(v_i(t))+d_{b,i}(t)+W_i.
\end{aligned}
\end{equation}
Thus, vehicle~$i$ keeps the spacing policy $h(v_i)+d_{b,i}$
within a tube of width $W_i$.
Recall from Step~5 that
$h^{-1}:[0,\infty)\to[0,\vmax)$ is well-defined.
Since $h'\ge\lambda>0$, its inverse is increasing and satisfies
$h^{-1}(y+z)\le h^{-1}(y)+z/\lambda$
for $y,z\ge0$.
Furthermore, $0\le\sigma_a(r)\le\max\{0,r\}$ with
$\sigma_a(r)=0$ for $r\le0$, hence
$d_{b,i}(t)\ge0$ and $d_{b,i}(t)=0$ whenever
$v_i(t)\le v_{i-1}(t)$.
Moreover, we use that $\dot q_i=v_{i-1}-v_i$ in both modes
and that $v_i(t)\in[0,\vmax)$ by Step~3.
We show the two bounds in~\eqref{eq:linkwise-thm} separately.

\emph{Lower bound.}
Fix $t_1\ge0$. If $h(n_i(t_1))\le W_i$, then
\[
h^{-1}\Big(\max\big\{h(n_i(t_1))-W_i,0\big\}\Big)
=h^{-1}(0)=0\le v_i(t_1).
\]
Put $V:=n_i(t_1)$, let $h(V)>W_i$, and suppose that
\[
v_i(t_1)\le h^{-1}(h(V)-W_i)<V.
\]
Since $v_i(0)\ge V$ and $v_i$ is continuous,
\[
t_0:=\max\{s\in[0,t_1]:v_i(s)\ge V\}
\]
exists, $t_0<t_1$, and
$v_i(s)<V\le v_{i-1}(s)$ for all $s\in(t_0,t_1]$.
Hence $\dot q_i>0$ on $(t_0,t_1]$, which gives
$q_i(t_0)<q_i(t_1)$, and $d_{b,i}=0$ on $[t_0,t_1]$.
Thus,~\eqref{eq:tube} implies
$q_i(t_0)>h(v_i(t_0))=h(V)$, since
$v_i(t_0)=V$ by continuity and the definition of $t_0$,
and
$q_i(t_1)<h(v_i(t_1))+W_i\le h(V)$
by the assumed bound on $v_i(t_1)$ and the monotonicity of $h$.
This yields
\[
h(V)<q_i(t_0)<q_i(t_1)<h(V),
\]
a contradiction.

\emph{Upper bound.}
Fix $t_1\ge0$ and put $Z:=N_i(t_1)$.
If $v_i(t_1)\le Z$, the upper bound follows immediately
from $W_i>0$ and the strict monotonicity of $h^{-1}$.
Otherwise, suppose that $v_i(t_1)>Z$.

Since $v_i(0)\le Z$,
\[
t_0:=\max\{s\in[0,t_1]:v_i(s)\le Z\}
\]
exists, $t_0<t_1$, and
$v_i(s)>Z\ge v_{i-1}(s)$ for all $s\in(t_0,t_1]$.
Hence $\dot q_i<0$ on $(t_0,t_1]$, which gives
$q_i(t_0)>q_i(t_1)$, and
$d_{b,i}(t_0)\ge0$ as well as
$d_{b,i}(t_1) > 0$.
Thus, by~\eqref{eq:tube},
$q_i(t_1)>h(v_i(t_1))$.
Using $v_i(t_0)=Z$, $v_{i-1}(t_0)\ge0$, and
$\sigma_a(r)\le\max\{0,r\}$ in~\eqref{eq:db}, we obtain
\begin{multline*}
q_i(t_0)
<h(v_i(t_0))
+\sigma_a\Big(
\frac{v_i(t_0)^2-v_{i-1}(t_0)^2}{2\delta}
\Big)+W_i\\
\le h(Z)+\frac{Z^2}{2\delta}+W_i.
\end{multline*}
Therefore,
$h(v_i(t_1))<h(Z)+W_i+Z^2/(2\delta)$,
which proves~\eqref{eq:linkwise-thm}.

Finally, since
\[
h^{-1}(\max\{y-W_i,0\})
\ge h^{-1}(y)-W_i/\lambda
\]
and
$h^{-1}(y+z)\le h^{-1}(y)+z/\lambda$
for $y,z\ge0$, and since $N_i(t)<\vmax$,
\eqref{eq:linkwise-thm} implies
\[
n_i(t)-\frac{W}{\lambda}
\le v_i(t)
<N_i(t)+\frac{1}{\lambda}
\Big(W+\frac{\vmax^2}{2\delta}\Big).
\]
Thus, \eqref{eq:string-stab} holds with
$\Theta=(W+\vmax^2/(2\delta))/\lambda$,
which is independent of $i$ and $N$.
This proves assertion~(iv) and completes the proof.
\end{proof}

\bibliographystyle{IEEEtran}
\bibliography{references}
\begin{IEEEbiography}[{\includegraphics[width=1.0in,height=5.95in,clip,keepaspectratio]{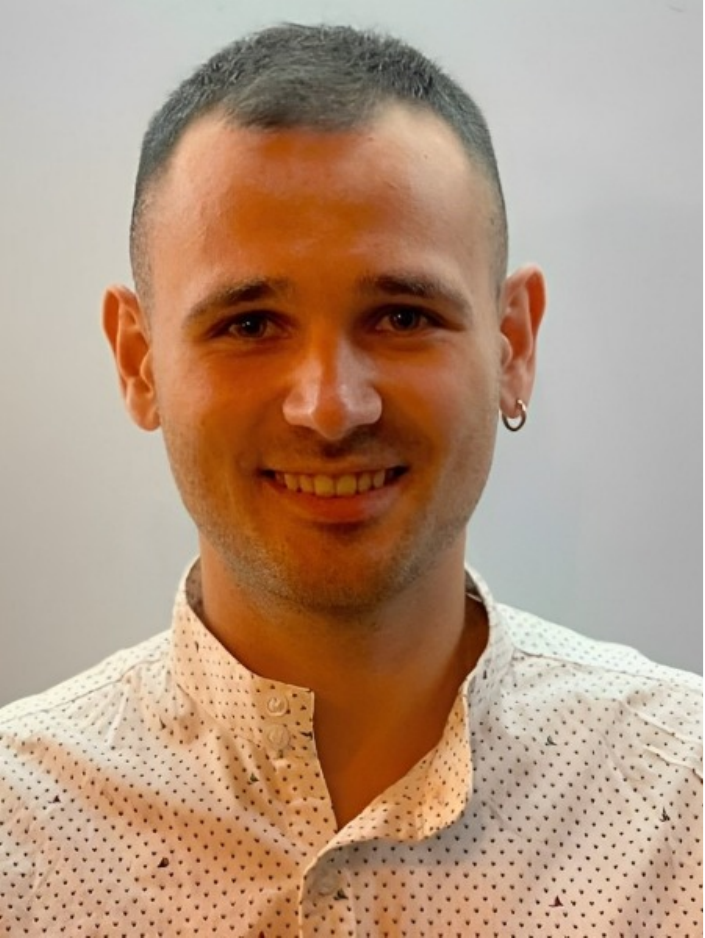}}]
{Panagiotis S. Trakas} was born in Tripoli, Greece, in 1998. He received the Diploma in Electrical and Computer Engineering from the Aristotle University of Thessaloniki, Thessaloniki, Greece, in 2021 and the Ph.D. degree in Electrical and Computer Engineering from the University of Patras, Patras, Greece in 2025. 

He is currently a Postdoctoral Researcher at the Automatic Control Laboratory, Department of Information Technology and Electrical Engineering, ETH Zurich, Switzerland. His research interests include nonlinear robust adaptive control, learning-based control, and data-driven uncertainty quantification and safety assurance for autonomous systems.
\end{IEEEbiography}

\begin{IEEEbiography}[{\includegraphics[width=1in,height=1.25in,clip,keepaspectratio]{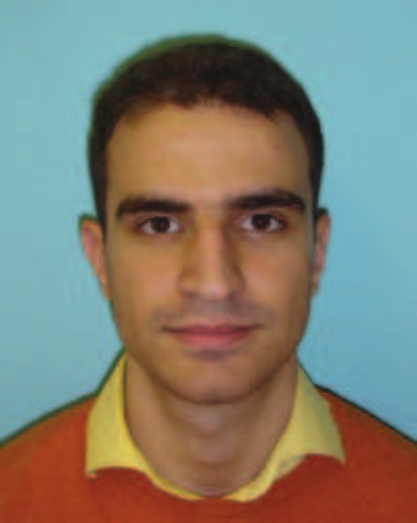}}]
{Charalampos P. Bechlioulis}(Senior Member, IEEE) was born in Arta, Greece, in 1983. He received the diploma degree (first in his class) in electrical and computer engineering (2006), the bachelor of science degree (second in his class) in mathematics (2011), and the Ph.D. degree in electrical and computer engineering (2011) from the Aristotle University of Thessaloniki, Thessaloniki, Greece.

He is currently a Professor with the Division of Systems and Control, Department of Electrical and Computer Engineering, University of Patras, Patras, Greece. He has authored more than 140 papers in scientific journals and conference proceedings and three book chapters. His research interests include nonlinear control with prescribed performance, system identification, control of robotic vehicles, and multiagent systems.
\end{IEEEbiography}

\begin{IEEEbiography}[{\includegraphics[width=1in,height=1.25in,clip,keepaspectratio]{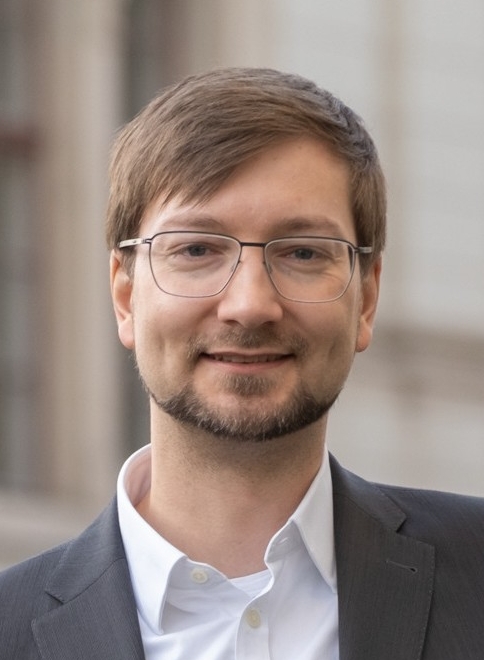}}]{Thomas Berger} was born in Germany in 1986. He received his Ph.D. in Mathematics from Technische Universität Ilmenau, Germany, in 2013. From 2013 to 2018 Dr. Berger was a postdoctoral researcher at Universität Hamburg, Germany, and from 2019 to 2025 he was a Juniorprofessor at Universität Paderborn, Germany. Since January 2026 Dr. Berger is a Heisenberg-Professor at the  Martin-Luther-Universität Halle-Wittenberg, Germany. His research interest encompasses adaptive control, optimization-based control, differential–algebraic systems and multibody dynamics.

For his exceptional scientific achievements in the field of Applied Mathematics and Mechanics, he received the ``Richard-von-Mises Prize 2021'' of the International Association of Applied Mathematics and Mechanics (GAMM). Dr. Berger further received several awards for his dissertation, including the ``2015 European Ph.D. Award on Control for Complex and Heterogeneous Systems'' from the European Embedded Control Institute and the ``Dr.-Körper-Preis 2015'' from the GAMM. He serves as an Associate Editor for Mathematics of Control, Signals, and Systems, and the IMA Journal of Mathematical Control and Information.
\end{IEEEbiography}

\end{document}